\documentclass{aa}  

\usepackage{graphicx}
\usepackage{txfonts}
\usepackage{xspace}
\usepackage{multirow}
\newcommand{\Msun}{$M_{\odot}$\xspace}

\newcommand{\vlsr}{$\varv_{\rm{LSR}}$\xspace}

\newcommand{\kms}{km\,s$^{-1}$\xspace}

\newcommand{\fCO}{$f_{\rm{CO}}$}

\newcommand{\COone}{$^{12}\rm{CO}\,(1-0)$\xspace}
\newcommand{\COtwo}{$^{12}\rm{CO}\,(2-1)$\xspace}
\newcommand{\COthree}{$^{12}\rm{CO}\,(3-2)$\xspace}
\newcommand{\COfour}{$^{12}\rm{CO}\,(4-3)$\xspace}

\newcommand{\Htwo}{H$_2$\xspace}

\newcommand\Tshell{$T_{\rm{sh}}$}

\newcommand\Mshellg{$M_{\rm{sh,g}}$}

\newcommand\Rshell{$R_{\rm{sh}}$}
\newcommand\dRshell{$\Delta R_{\rm{sh}}$}
\newcommand{\vshell}{$\varv_{\rm{sh}}$\xspace}

\xspace

\xspace
\xspace

\xspace

\begin{document}

\title{Probing the dynamical and kinematical structures of detached shells around AGB stars}

  \author{M.~Maercker \inst{1}
        \and
          E.~De Beck \inst{1}
          \and          
          T.~Khouri \inst{1}
          \and
          W.H.T.~Vlemmings \inst{1}
                  \and
            J.~Gustafsson \inst{1}
          \and
          H.~Olofsson \inst{1}
           \and
          D.~Tafoya \inst{2}
          \and
          F.~Kerschbaum \inst{3}
          \and 
          M.~Lindqvist \inst{2}
}

     \institute{Department of Space, Earth and Environment, Chalmers University of Technology, 41296 Gothenburg, Sweden\\
   \email{maercker@chalmers.se}
         \and
         Department of Space, Earth and Environment, Chalmers University of Technology, Onsala Space Observatory, 43992 Onsala, Sweden
         \and
             Department of Astrophysics, University of Vienna, T\"urkenschanzstr. 17, 1180 Vienna, Austria\\
             }

   \date{Received February 17, 2024; accepted April 26, 2024}

    \abstract
   {The chemical evolution of asymptotic giant branch (AGB) stars is driven by repeated thermal pulses (TPs). The duration of a TP is only a few hundred years, whereas an inter-pulse period lasts $10^4-10^5$ years. Direct observations of TPs are hence unlikely. However, the detached shells seen in CO line emission that are formed as a result of a TP provide indirect constraints on the changes experienced by the star during the pulse.}
   {We aim to resolve the spatial and kinematic sub-structures in five detached-shell sources to provide detailed constraints for hydrodynamic models that describe the formation and evolution of the shells.}
   {We use observations of the \COone emission towards five carbon-AGB stars with ALMA (Atacama Large Millimeter/submillimeter Array), including previously published observations of the carbon AGB star U~Ant. The data have angular resolutions of 0\farcs3 to 1\arcsec and a velocity resolution of 0.3\,\kms. This enables us to quantify spatial and kinematic structures in the shells. Combining the ALMA data with single-dish observations of the \COone to \COfour emission towards the sources, we use radiative transfer models to compare the observed structures with previous estimates of the shell masses and temperatures.}
   {The observed emission is separated into two distinct components: a more coherent, bright outer shell and a more filamentary, fainter inner shell. The kinematic information shows that the inner sub-shells move at a higher velocity relative to the outer sub-shells. The observed sub-structures reveal a negative velocity gradient outwards across the detached shells, confirming the predictions from hydrodynamical models. However, the models do not predict a double-shell structure, and the CO emission likely only traces the inner and outer edges of the shell, implying a lack of CO in the middle layers of the detached shell. Previous estimates of the masses and temperatures are consistent with originating mainly from the brighter subshell, but the total shell masses are likely lower limits. Also, additional structures in the form of partial shells outside the detached shell around V644~Sco, arcs within the shell of R~Scl, and a partially filled shell for DR~Ser indicate a more complicated evolution of the shells and mass-loss process throughout the TP-cycle than previously assumed.}
  {The observed spatial and kinematical splittings of the shells appear consistent with results from hydrodynamical models, provided the CO emission does not trace the \Htwo density distribution in the shell but rather traces the edges of the shells. The hydrodynamical models predict very different density profiles depending on the evolution of the shells and the different physical processes involved in the wind-wind interaction (e.g. heating and cooling processes). It is therefore not possible to constrain the total shell mass based on the CO observations alone. Additional features outside and inside the shells complicate the interpretation of the data. Complementary observations of, e.g., CI as a dissociation product of CO would be necessary to understand the distribution of CO compared to \Htwo, in addition to new detailed hydrodynamical models of the pre-pulse, pulse, and post-pulse wind. Only a comprehensive combination of observations and models will allow us to constrain the evolution of the shells and the changes in the star during the thermal-pulse cycle.}
  

  
   \maketitle
%
\nolinenumbers
\section{Introduction}
Stars with main-sequence masses from 0.5\,\Msun up to $8-10$\,\Msun will spend the last few million years of their evolution on the asymptotic giant branch (AGB). On the AGB, the star consists of a dormant carbon-oxygen core, surrounded by thin helium- and hydrogen-burning shells above which there is a deep convective envelope that mixes material to the stellar surface~\citep[e.g.,][]{karakasco2007,karakasco2014}.

The chemical evolution of AGB stars is driven by thermal pulses (TPs) -- periods of explosive helium burning in the shell surrounding the stellar core. The restructuring of the hydrogen- and helium-burning shells during and after a TP, and the effect of subsequent TPs, leads to the nucleosynthesis of heavy elements~\citep{karakasco2014}. The newly synthesised elements are brought to the surface by convective currents in a process known as the third dredge-up. The stellar winds experienced by AGB stars include these elements and eventually release them to the interstellar medium (ISM). The process is repeated until the star ends its evolution on the thermally pulsing AGB. 

The helium burning during a TP lasts only a few hundred years. The inter-pulse period, characterised by shell-hydrogen burning, depends on the stellar mass and can last between $10^4-10^5$ years~\citep{karakasco2014}. 
The mass loss during the inter-pulse periods limits the total lifetime on the AGB and the number of TPs a star can experience. Through TPs and the subsequent mass-loss on the AGB, low- to intermediate mass stars are major contributors to the chemical evolution of galaxies and the universe~\citep[e.g.,][]{bussoetal1999,cherchneff2006}, and determining the properties of the TP-cycle is important for our understanding of the yields from AGB stars to galaxies.

The increase in surface luminosity combined with less tightly bound surface layers of the star during a TP cause an increase in the stellar mass loss by up to two orders of magnitude and an increase in wind expansion velocity by $\sim$10 \kms~\citep[e.g.,][]{mattssonetal2007}. After the TP, both the mass-loss rate and expansion velocity decrease again to their pre-pulse values, resulting in a thin shell of dust and gas that appears detached from the subsequent stellar wind (that is, mostly lacking circumstellar material inside the shell). Detached shells observed in CO emission around carbon AGB stars have been linked to this predicted increase in mass-loss rate and expansion velocity during a thermal pulse~\citep{olofssonetal1990, olofssonetal1993a,olofssonetal1993b, olofssonetal1996, schoieretal2005, maerckeretal2012,maerckeretal2016a,kerschbaumetal2017}. While no detached CO-shells have been detected around M-type or S-type stars, detached dust shells have been observed around AGB and post-AGB sources with varying chemistries~\citep[e.g.,][]{watersetal1994,izumiuraetal1996,izumiuraetal1997,hashimotoetal1998,specketal2000}. {However, these are most likely of a different origin.}

In general, the mass loss from AGB stars is assumed to be driven through radiation pressure on dust grains and dust-gas interaction~\citep{hofnerco2018}. The co-spatiality of the dust and gas indicates a common evolution of the two components and a possible connection to the mass-loss properties during the TP. The observed structures have also been reproduced successfully using one-dimensional hydrodynamic models, showing the potential of detached shells for constraining the TP cycle~\citep{steffenco2000,mattssonetal2007}. However, spatial resolution and sensitivity limit the conclusions that can be drawn on the TP cycle from observations.

These uncertainties in the properties of the star during the TP, and the evolution of the stellar wind throughout the TP-cycle, limit our understanding of stellar evolution and the yields of new elements from AGB stars to the ISM. Direct observations of TPs are unlikely owing to their short duration and long inter-pulse timescales. TPs, therefore, have to be constrained indirectly, and studying the detached shells remains the best way of understanding the TP cycle.

There are currently seven known carbon AGB stars with detached CO shells: U~Ant, U~Cam, TT~Cyg, R~Scl, V644~Sco, S~Sct, and DR~Ser~\citep{schoieretal2005}. All of the CO detached shells are also detected in thermal dust emission or dust-scattered light~\citep[e.g.,][]{delgadoetal2001,maerckeretal2010,maerckeretal2014,olofssonetal2010,ramstedtetal2011}, and the dust is indeed found to be co-spatial with the gas. The observations of CO and dust both show thin shells expanding away from the central stars. The shells have radii of $\sim10^{17}\rm{cm}$ (angular sizes between 7\arcsec\/ and 60\arcsec). Based on the data that was available, the widths of the shells were estimated to be $\sim10^{16}\rm{cm}$ (angular sizes of $\approx2-4$\arcsec -- roughly corresponding to the highest available resolution). Expansion velocities have been measured to be $12-23$\,\kms. The sizes and expansion velocities give dynamical ages of $\approx1000-8000$ years.

Recent observations of the detached shell around U~Ant with ALMA (Atacama Large Millimeter/submillimeter Array) at a spatial resolution of $\approx$1\farcs4 show that what was previously assumed to be one shell appears to be split in two shells in filamentary structures expanding at different velocities \citep{kerschbaumetal2017}. Kerschbaum et al. interpret the splitting of the shell as a consequence of the hydrodynamic evolution of the shell, creating two separate shells - a faster outer shell and a slower inner shell - as the result of forward and reverse shocks, respectively. While one-dimensional hydrodynamical models show a gradient in velocity across the shell, they do not produce a double-shell structure, complicating a direct comparison between models and observations and consequently also an interpretation of the data. 

In this paper we present new observations towards the carbon AGB stars R~Scl, V644~Sco, S~Sct, and DR~Ser taken with ALMA in \COone~emission at high angular resolution (0\farcs3~$-$~1\arcsec).  We expected the high-angular resolution data to probe in detail the spatial and kinematical properties of the shells, including a similar splitting of the shells in space and velocity as was seen for U~Ant, enabling us to constrain their evolution. While we indeed observe the expected splitting in the emission, we provide a new interpretation of the data with a faster inner shell and a slower outer shell. The analysis of the data also shows that previous measurements of the estimated shell masses and interpretations of the detailed evolution of the shells are not necessarily straightforward.

The new observations are described in Section~\ref{s:observations}. The spatial and kinematical structures, as well as radiative transfer models of the shells are presented in Sect.~\ref{s:results}. The implications of the observed structures for the interpretation of the evolution of the detached shells and their connection to TPs is discussed in Sect.~\ref{s:discussion}. Our conclusions with suggested future research are presented in Sect.~\ref{s:conclusions}.

\begin{table*}[ht]
\tiny
\centering
\caption{Observational parameters for the ALMA data. The observations for U~Ant were already presented in~\cite{kerschbaumetal2017}.}
\begin{tabular}{l c c c c c  l l}
\hline\hline\\[-2ex]
Source	&$\theta_{\rm{beam}}$ & PA$_{\rm{beam}}$ & MRS &$\Delta \varv$ & rms &Arrays & Project(s)\\
		&	[$\arcsec\times\arcsec$] & [deg] &	[\arcsec]&[\kms] &	[mJy beam $^{-1}$] & &\\
\hline \\[-2ex]
U~Ant	&1.37$\times$1.10& 86.6 &--&1.0&8.0&TP$^*$, ACA, TM &2015.1.00007.S\\ 
R~Scl	&0.80$\times$0.64&86.7&65.0&0.35&3.5&ACA, TM&2021.1.00313.S\\ 
V644~Sco& 0.51$\times$0.43 & -71.9&23.4 & 0.65 &2.0& TM&2018.1.00010.S, 2019.1.00021.S, 2021.1.00313.S\\ 
S~Sct	&1.50$\times$1.17&-67.8&76.3&0.65&3.5&ACA, TM&2017.1.00006.S, 2019.1.00021\\ 
DR~Ser	&0.52$\times$0.50&-29.0&24.9&0.64&1.8&TM&2018.1.00010.S, 2019.1.00021.S, 2021.1.00313.S\\ 
\hline
\end{tabular}%
  \label{t:almaobs}%
  
  \tablefoot{Table columns are: beam full width at half maximum and position angle ($\theta_{\rm{beam}}$ and PA$_{\rm{beam}}$), respectively, the maximum recoverable scale (MRS), the velocity resolution ($\Delta \varv$), and the rms noise. For each source the arrays (TP -- total power; ACA -- Atacama Compact Array; TM -- twelve metre array) and the original project IDs that were combined are given. Parameters are given for the final, combined data cubes. \\$^*$ The TP data for U~Ant were combined with the ACA and TM arrays through feathering, leaving possible artefacts caused by resolved-out flux. Details are of the data handling for U~Ant are explained in~\cite{kerschbaumetal2017}.
}
\end{table*}%

\begin{table*}[t]
\centering
\caption{Peak intensities of ALMA spectra (extracted from the ALMA data centred on the star from apertures with diameters corresponding to $\approx$10\% of the shell diameter) and single-dish observations of CO towards the detached-shell sources. }
\begin{tabular}{l c c c c c}
\hline\hline\\[-2ex]
 & \multicolumn{1}{c}{Telescope}& \multicolumn{1}{c}{\COone} & \multicolumn{1}{c}{\COtwo} & \multicolumn{1}{c}{\COthree} & \multicolumn{1}{c}{\COfour} \\

 \hline
U~Ant  &ALMA & & 142& &\\
&SEST  &0.48 &0.89  &0.93 &\\
R~Scl  && \multicolumn{3}{l}{not modelled in this paper} &\\
V644~Sco &ALMA & 12.6  & &  &\\
&SEST & 0.19 &0.75 &&\\
&APEX & &&1.2 &1.25\\
S~Sct  &ALMA &25 & & &\\
  &SEST & 0.87  &0.64  &0.16  &\\
    &APEX &    &            &  0.3   & 0.02\\
DR~Ser &ALMA &14            &            &  &\\
 &APEX &            &            & 0.41  & 0.28\\
\hline \\[-2ex]
Beam & SEST & 44\arcsec (0.75) & 23\arcsec (0.50) & 15\arcsec (0.25) & \\
 & APEX & & & 17\arcsec (0.79) & 13\arcsec (0.58)\\
 \hline
\end{tabular}
\label{t:sdobs}
 \tablefoot{Intensities are in mJy for the ALMA spectra and in K (main-beam temperature, $T_{\rm{mb}}$) for the single-dish spectra. SEST observations are taken from \cite{schoieretal2005} and references therein. The APEX observations are new. We assume an uncertainty of 20\% in the peak intensities. The full width half maximum of the telescope beams are given for the respective transitions in the last two rows, with the main-beam efficiencies in parentheses.
}
\end{table*}

\section{Observations}
\label{s:observations}

\subsection{Atacama Large Millimeter/submillimeter Array}
\label{s:alma}
We observed the detached shells around R~Scl, V644~Sco, S~Sct, and DR~Ser with ALMA in several observing programs (see Table~\ref{t:almaobs} for the program IDs). The raw data of the observations were retrieved from the ALMA Science Archive. The calibrated measurement sets were derived by applying calibration tables, generated by the ALMA pipeline during the QA2 phase, to the raw data using the corresponding versions of both the ALMA pipeline and CASA~\citep{CASAref}. Following this, data combination and imaging tasks were carried out with \textsc{CASA} version 6.4. The data were merged using the task `\textit{concat}'. During this process, the visibility weights of the different arrays were adjusted for uniformity. Subsequently, spectral cubes were generated using the \textsc{CASA} task `\textit{tclean}'. Here, the `\textit{deconvolver}' parameter was set to `\textit{multiscale}' mode. For imaging, the Briggs weighting scheme was employed, with the `\textit{robust}' parameter set to values between 0.5 and 0.8, optimising the balance between sensitivity and resolution. For the sources R~Scl and S~Sct, imaging was gridded in mosaic mode. The properties of the final image cubes of the ALMA observations are given in Table~\ref{t:almaobs}. The scripts that were used for the calibration and imaging are available upon request.

The new ALMA data was complemented with already published data of U~Ant~\citep{kerschbaumetal2017}.

\subsection{Single-dish observations}
\label{s:apex}

Modelling multiple transitions of CO is an effective way of constraining the masses and temperatures in the detached shells~\citep[e.g.][]{schoieretal2005,maerckeretal2016a}. 
To this end, we use archival observations of CO towards detached shell sources~\citep[][and references therein]{schoieretal2005}, complemented with new observations using APEX \citep[Atacama Pathfinder Experiment;][]{gusten2006_apex} to cover the transitions from \COone~ up to \COfour. The complementary APEX observations cover the \COthree\/ and \COfour\/ transitions towards DR~Ser, V644~Sco, and S~Sct.  A summary of all the observations is presented in Table~\ref{t:sdobs}.


\COthree\/ observations were carried out with the SEPIA345 receiver \citep{sepiameledin2022} on 9 and 10 October 2022, with precipitable water vapour (PWV) values of $0.7-0.8$\,mm. \COfour\/ observations were carried out with the nFLASH460 receiver on 4, 6, and 7 October 2022, at PWV of $0.4-0.8$\,mm.  The fully calibrated data, delivered in corrected antenna temperature $T_{\rm{A}}^{*}$, were post-processed only through averaging of the separate spectra and subtracting a first-degree polynomial baseline fitted to the emission-free parts of the spectrum. The final spectra have characteristic rms noise levels of $11-16$\,mK at a velocity resolution of 0.9\,km\,s$^{-1}$. Lastly, we corrected the spectra for the relevant main-beam efficiencies\footnote{\tiny\texttt{https://www.apex-telescope.org/telescope/efficiency/}}.

{{The beam full width at half maximum (FWHM) for the SEST and APEX observations and the main-beam efficiencies are given for all transitions in Table~\ref{t:sdobs}.}}



\section{Results}
\label{s:results}

\begin{table*}[t]
\centering
\tiny
\caption{Derived parameters for the detached shells.}
    \begin{tabular}{lcccccccccccc}
\hline\hline\\[-2ex]
Source	& Distance  & \vlsr&  Shell	&\Rshell	& \dRshell & pa range &\vshell&\Mshellg &  \Tshell &\multicolumn{2}{c}{Separation} & Age\\
		&[pc] & [\kms]&		&[\arcsec]	& [\arcsec] &[deg]&[\kms]&$[10^{-3}$\Msun] & [K] &spatial&velocity & [yr]\\
\hline \\[-2ex]
    U~Ant      & 260      &   24&  inner  & 41.2  & 1.5   & 350-352&21.0   & -- &  \multirow{ 2}{*}{200} & \multirow{ 2}{*}{2\farcs7  (6\%)}       &  \multirow{ 2}{*}{2.0 (10\%)}   &   \multirow{ 2}{*}{2600 - 2800 }\\
                    &                    &&  outer  & 43.8  & 1.7   & &19 .0  &1.9 &  &     &      &  \\
                    &       &             &&       &       &       &       & &     & & &  \\
    R~Scl      & 360     & -19& inner & 16.3  &  1.7  &225-227    & 16.5 &\multirow{ 2}{*}{--}  &\multirow{ 2}{*}{--} &\multirow{ 2}{*}{ 2\farcs0 (11\%)} & \multirow{ 2}{*}{2.0 (14\%)} &  \multirow{ 2}{*}{ 1700 - 1900 }\\
               &                   && outer & 18.3  &  1.1      && 14.5   &&  &       &       &  \\
          &       &            &&       &       &       &       &    &   &  & &\\
    V644~Sco   & 700   &-21.5     &  inner  & 9.0     & 0.5   &315-317  & 23.8 &-- &\multirow{ 2}{*}{170} &  \multirow{ 2}{*}{0\farcs2   (2\%) }   & \multirow{ 2}{*}{ 2.2 (10\%)}  &   \multirow{ 2}{*}{1300 - 1400 }\\
               &                    &&  outer  & 9.2   & 1.3   & &21.6  &2.5 &&        &       &  \\
          &       &             &&       &       &       &       &       & & &&\\
    S~Sct      & 385      &15  &  inner  & 63.3  & 2.2   &240-242& 17.9  &-- & \multirow{ 2}{*}{60}&  \multirow{ 2}{*}{7\farcs0 (10\%) }    & \multirow{ 2}{*}{ 1.6 (9\%)} &   \multirow{ 2}{*}{7200 - 7900 }\\
               &                  & &  outer  & 70.3  & 2.5     && 16.3  &7.5 &&       &         &  \\
          &       &            & &       &       &       &       &   &    & & & \\
    DR~Ser     & 700    &9  &  inner  & 6.3   & 0.7   &265-267& 22.1 &--  & \multirow{ 2}{*}{100} &  \multirow{ 2}{*}{0\farcs9 (14\%)  }   &  \multirow{ 2}{*}{ 2.6 (13\%)}  &   \multirow{ 2}{*}{1100 - 1200 }\\
               &              &     &  outer  & 7.3   & 0.8   && 19.5  &1.0 &&       &       &  \\
\hline
    \end{tabular}%
  \label{t:shellresults}
  \tablefoot{ \Rshell~ and \dRshell~ are derived from Gaussian fits to the AARPs {at the stellar \vlsr. The pa range gives the range of position angels over which the AARP was formed to measure \dRshell (\Rshell was measured over all position angles; see text for details). }Values for \vshell~ are measured in the \COone~ spectra extracted from the ALMA observations (in \COtwo~for U~Ant). Distances, \Mshellg~ and \Tshell~are taken from \cite{schoieretal2005}, except for U~Ant where we find a better fit to the data using the value of \Mshellg~given in the table.}
\end{table*}%

In the following sections it is important to keep in mind that the observed CO emission, that is the brightness distribution, does not necessarily trace the gas density distribution of H$_2$ (the dominant component of the gas). Structures seen in the emission (see Sect.~\ref{s:kinematics}) hence do not directly translate into spatial (density) properties of the shells. As will be shown, the emission shows different structures, most prominently a splitting into two shells. Throughout the following sections we will refer to the structures seen in the CO emission. Whether this traces true density structures (e.g. two separate shells), is not clear and is discussed in Sect.~\ref{s:discussion}. 

\subsection{Kinematical and spatial structures}
\label{s:kinematics}

\begin{figure*}[t]
\centering
\includegraphics[width=4cm]{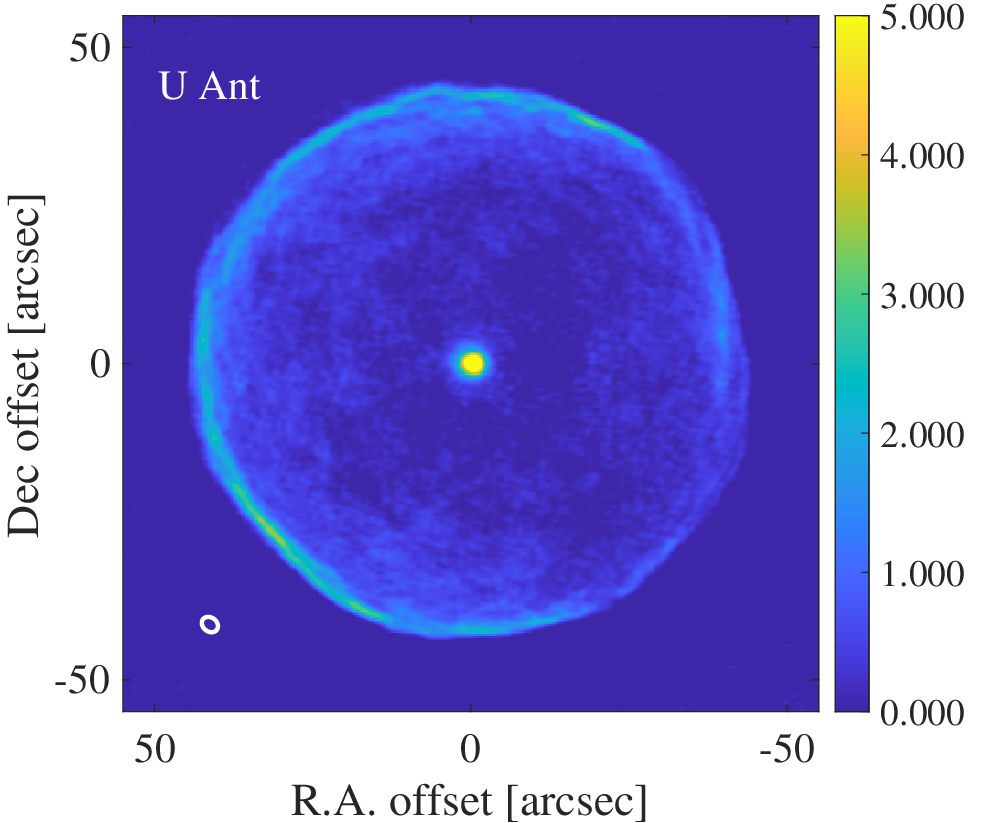} 
\includegraphics[width=3.25cm]{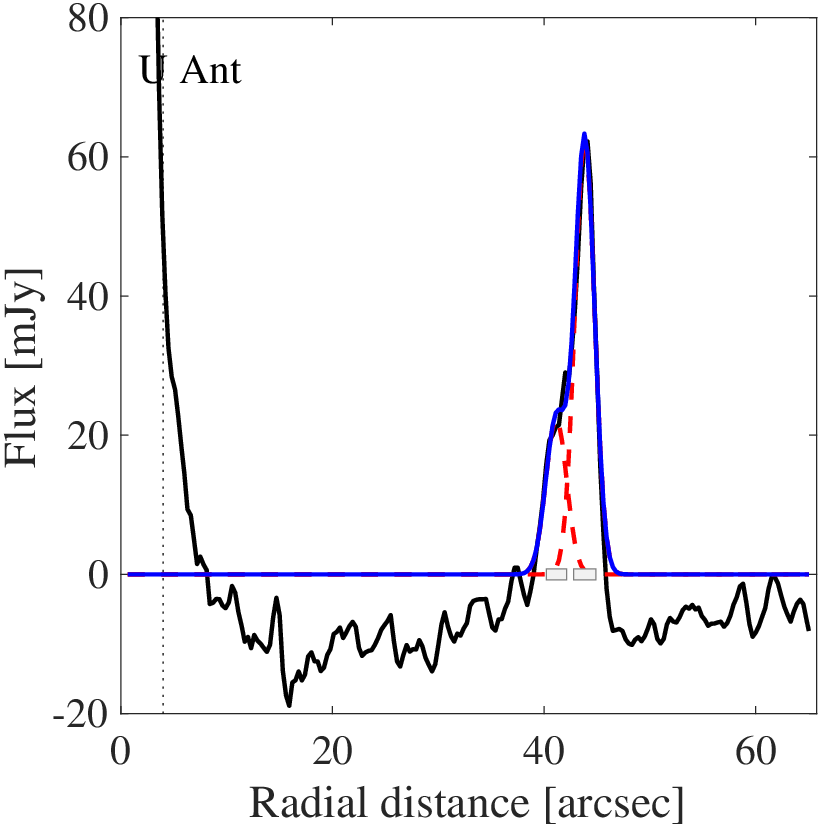} 
\includegraphics[width=4cm]{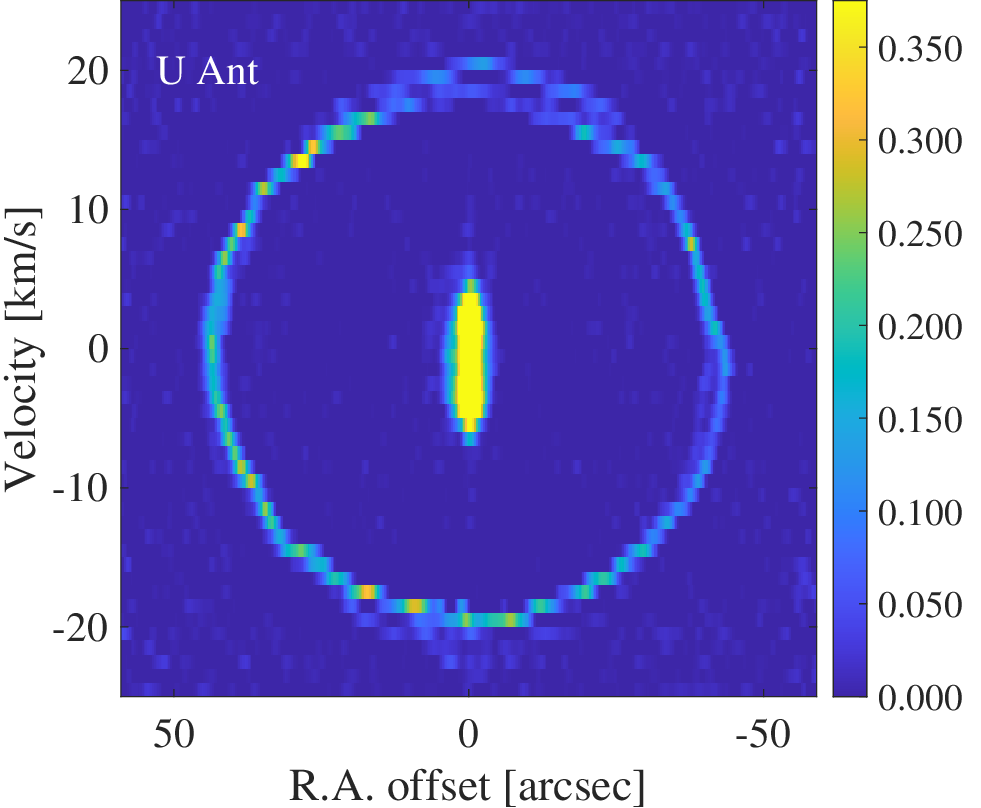}
\includegraphics[width=3.25cm]{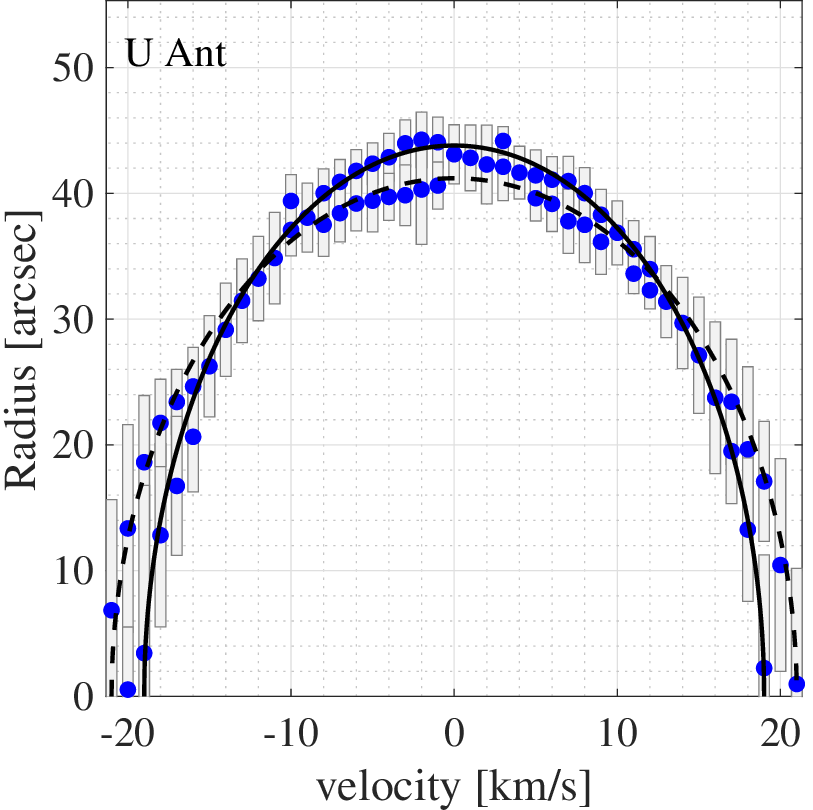}\\ 
\includegraphics[width=4cm]{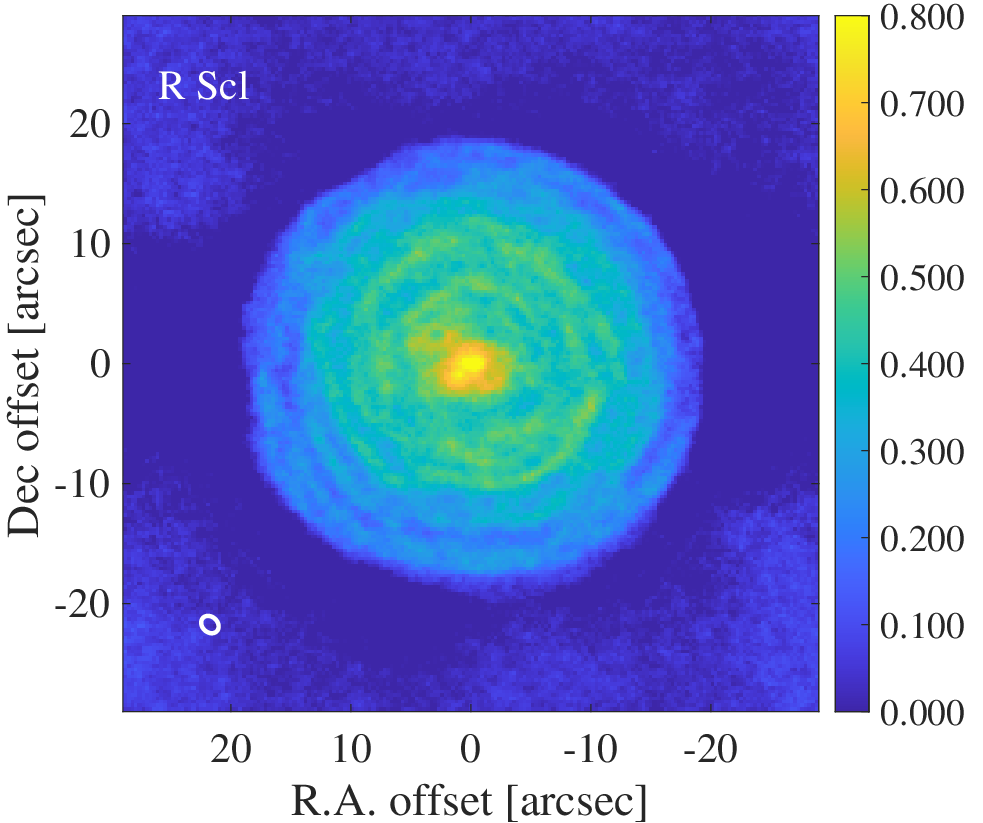} 
\includegraphics[width=3.25cm]{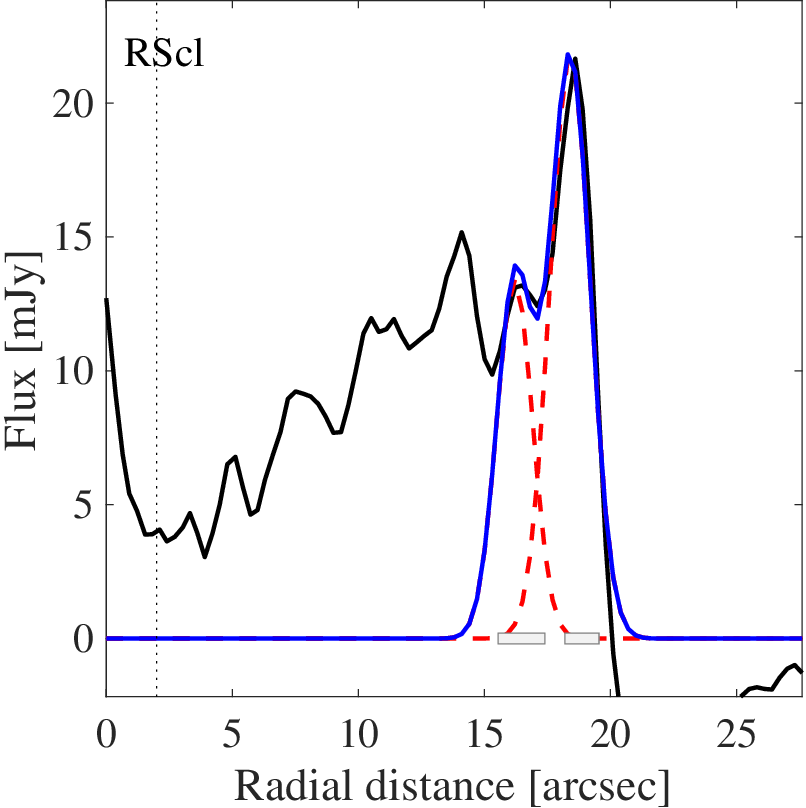} 
\includegraphics[width=4cm]{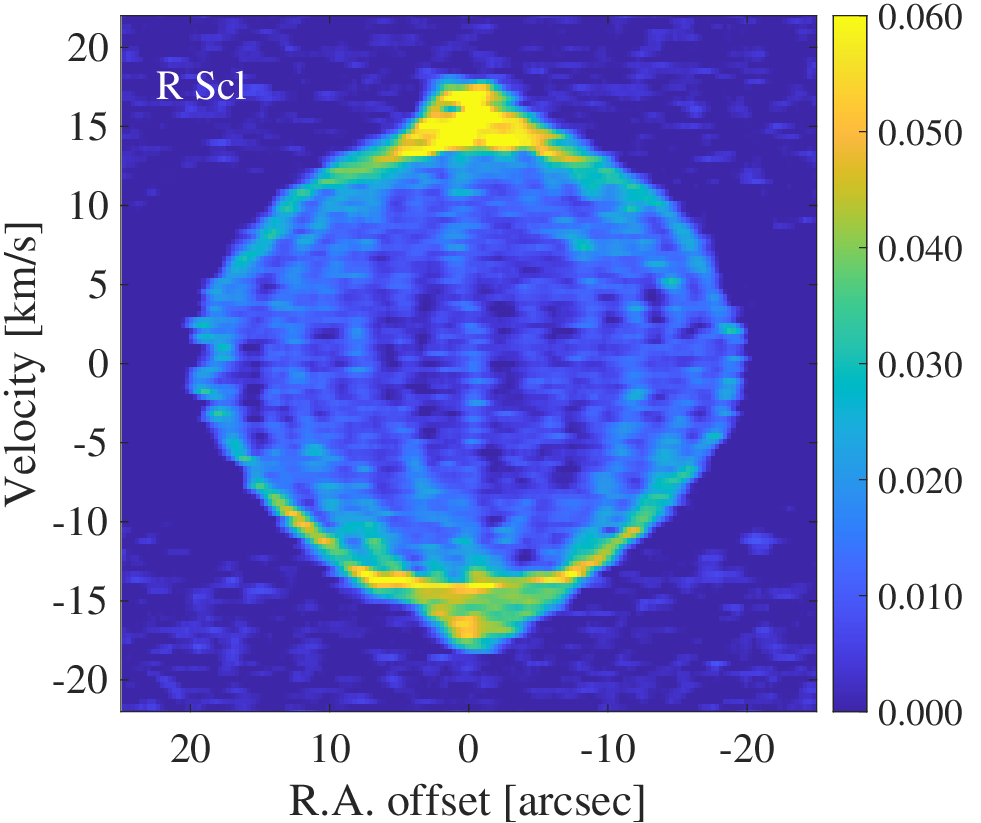} 
\includegraphics[width=3.25cm]{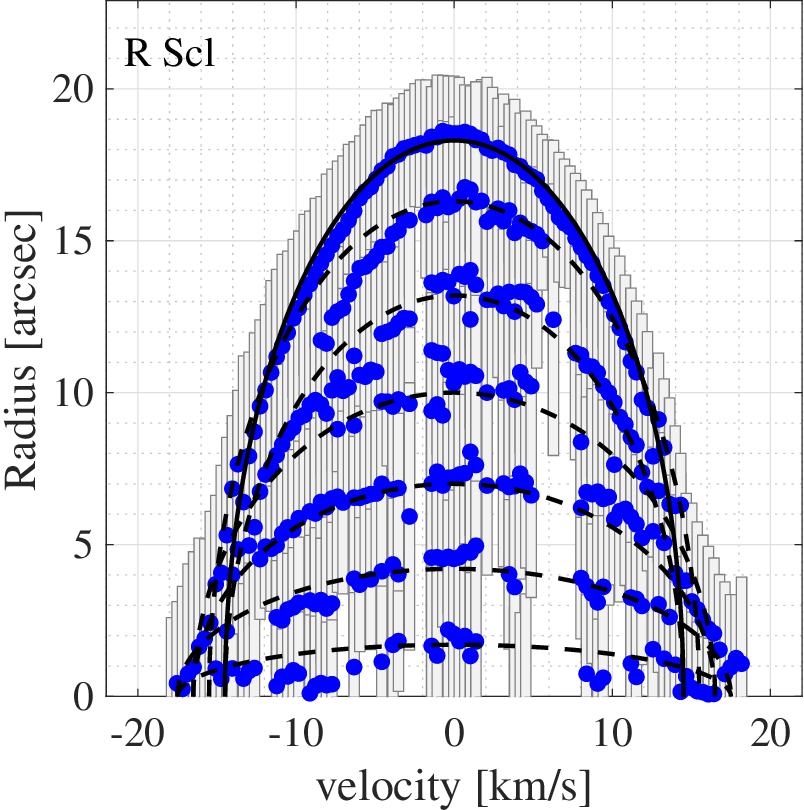}\\ 
\includegraphics[width=4cm]{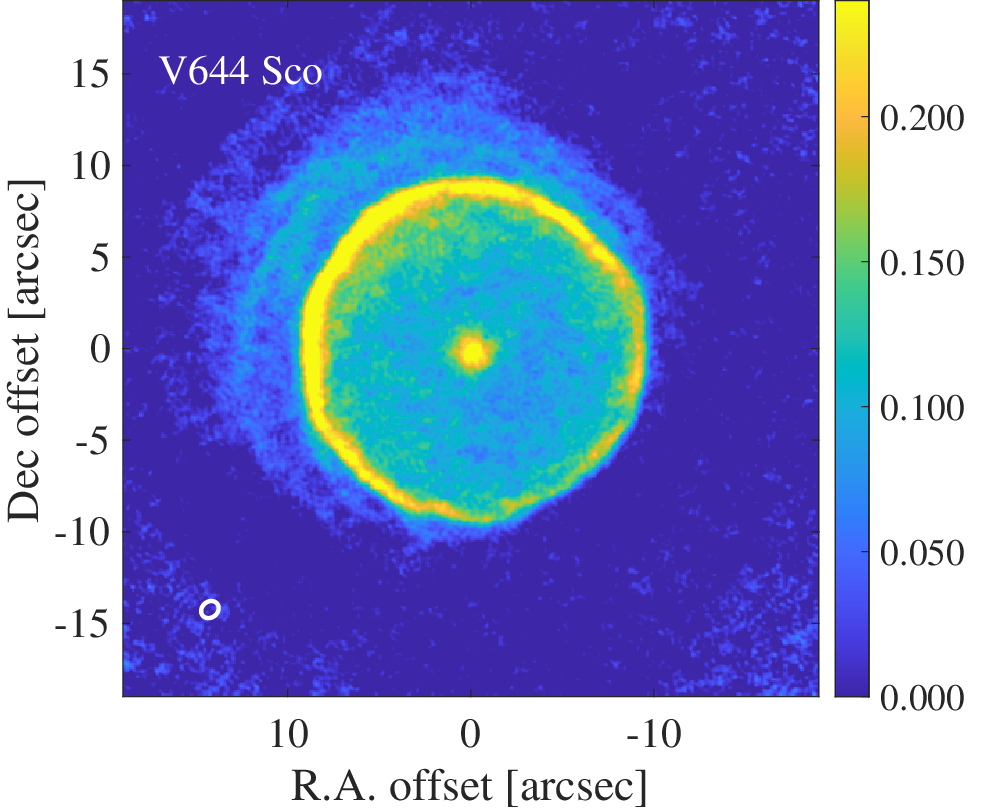} 
\includegraphics[width=3.25cm]{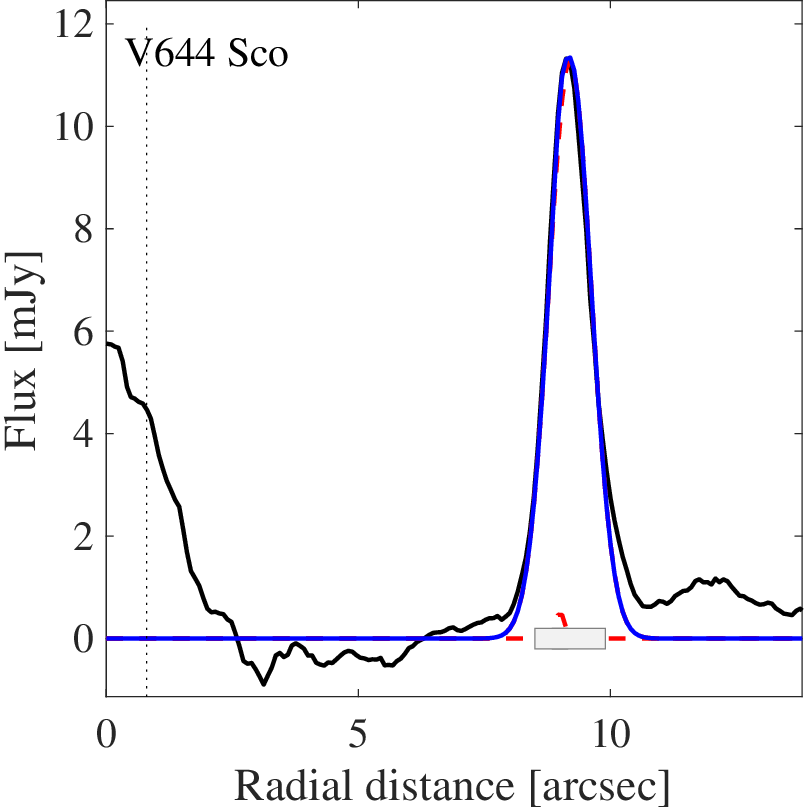} 
\includegraphics[width=4cm]{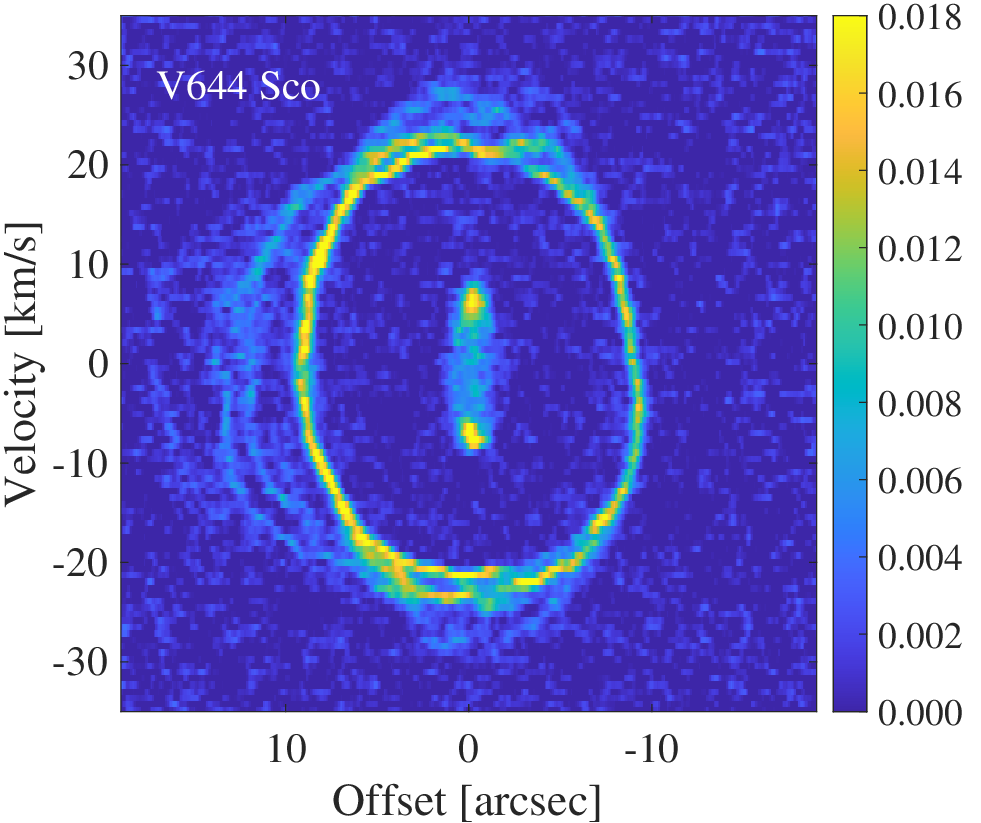}
\includegraphics[width=3.25cm]{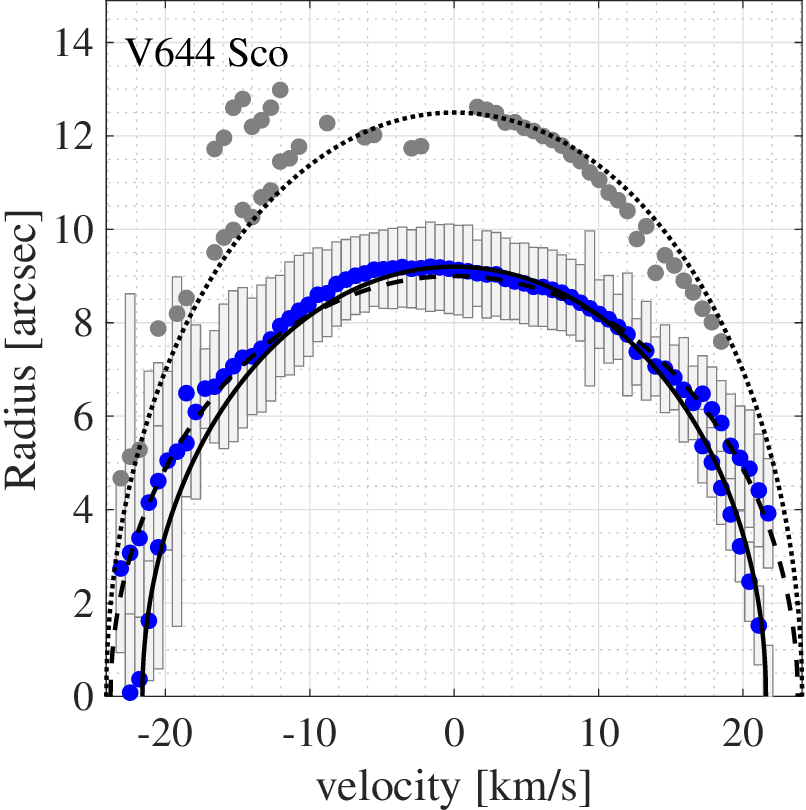}\\ 
\includegraphics[width=4cm]{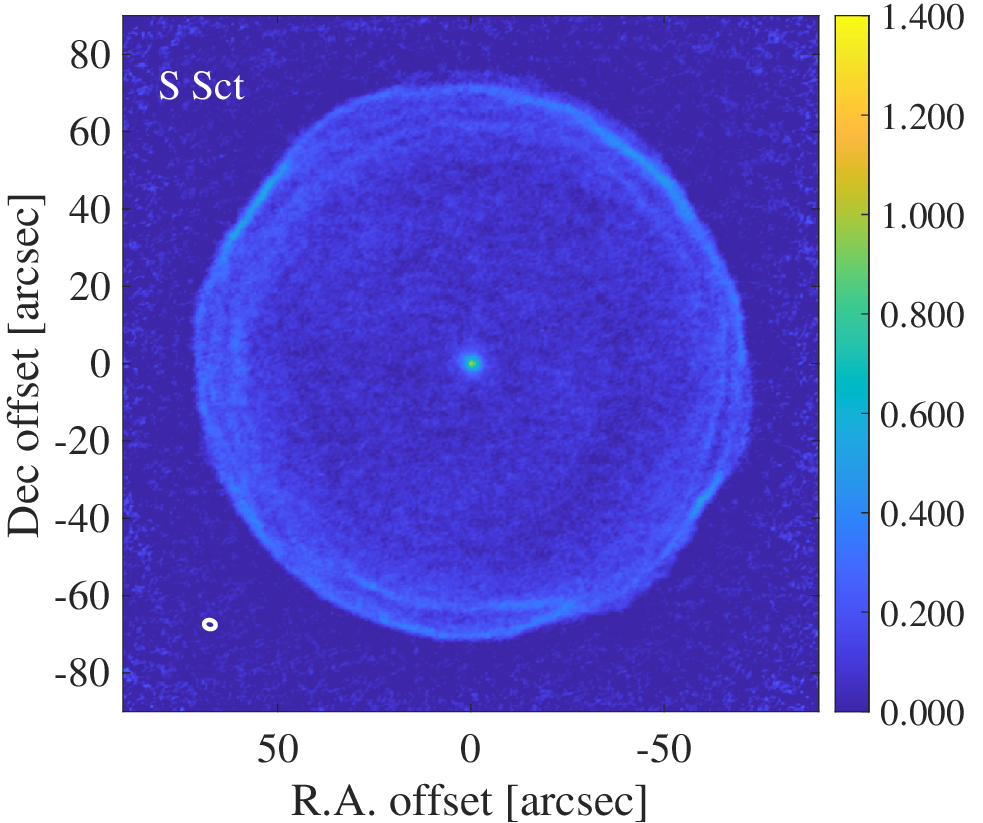} 
\includegraphics[width=3.25cm]{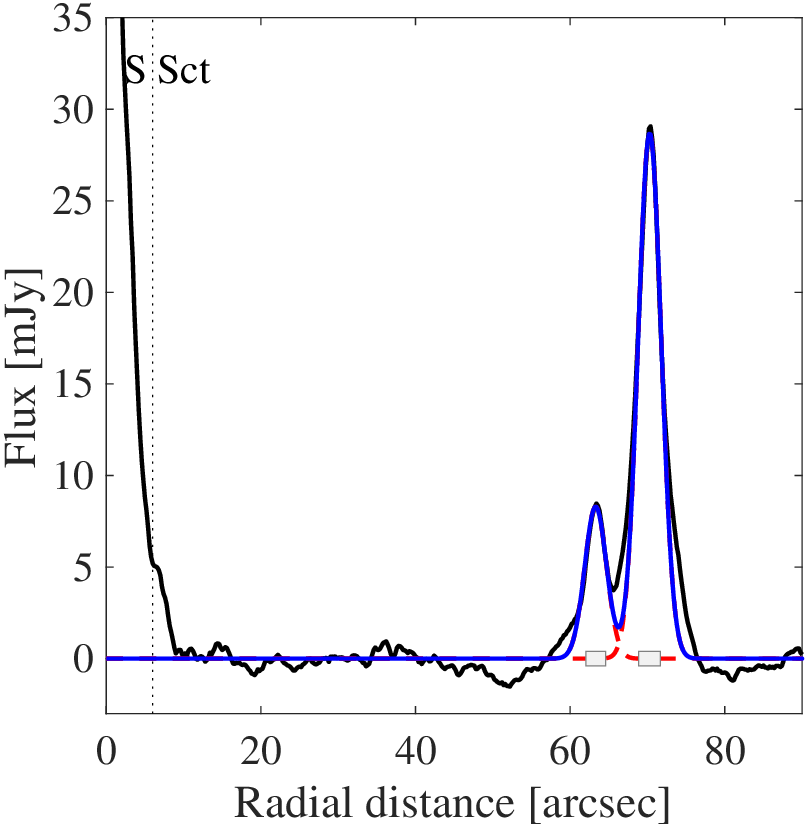} 
\includegraphics[width=4cm]{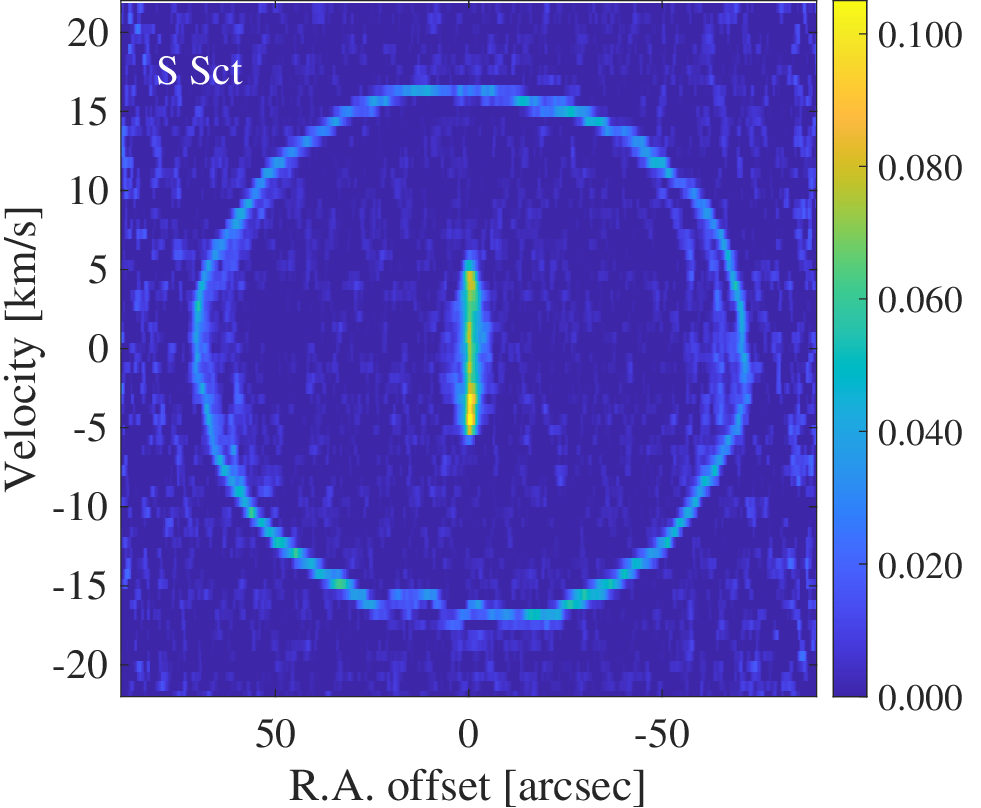}
\includegraphics[width=3.25cm]{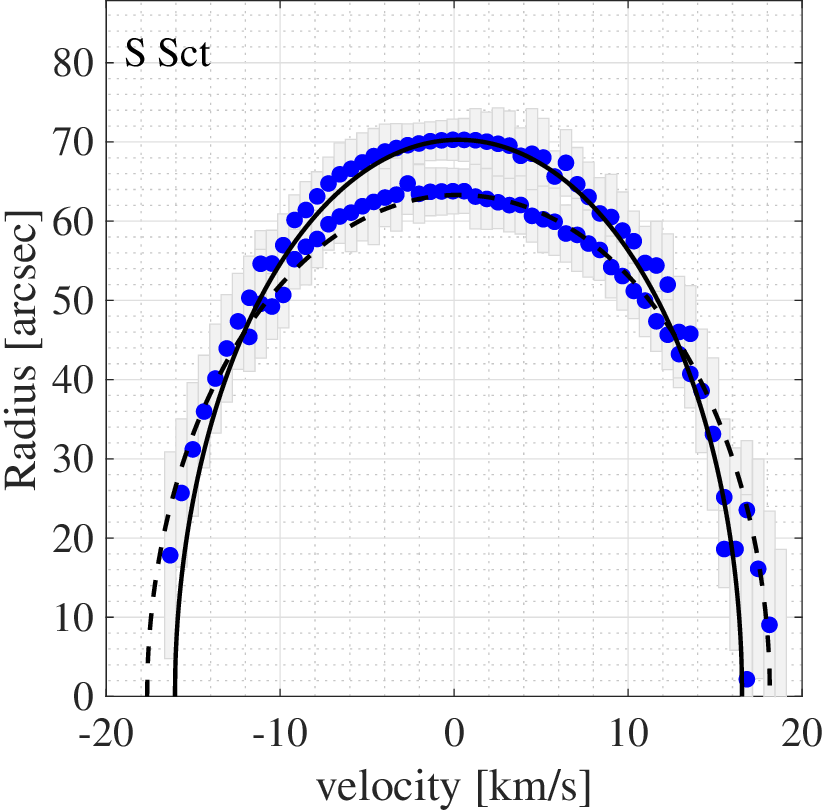}\\ 
\includegraphics[width=4cm]{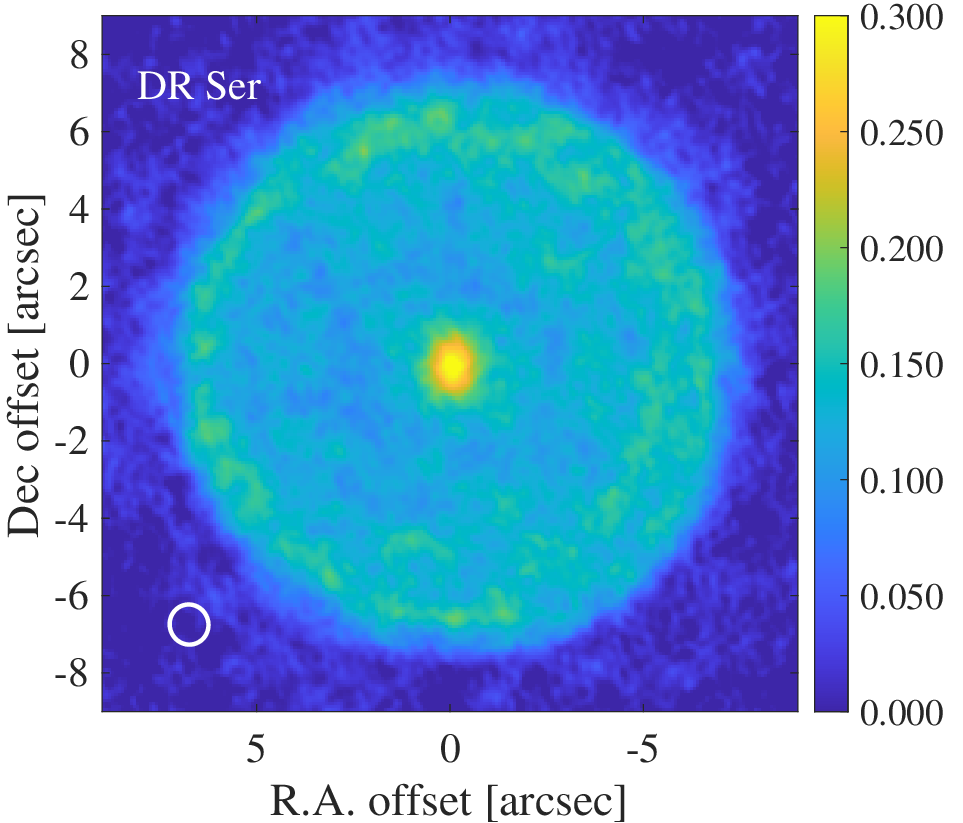} 
\includegraphics[width=3.25cm]{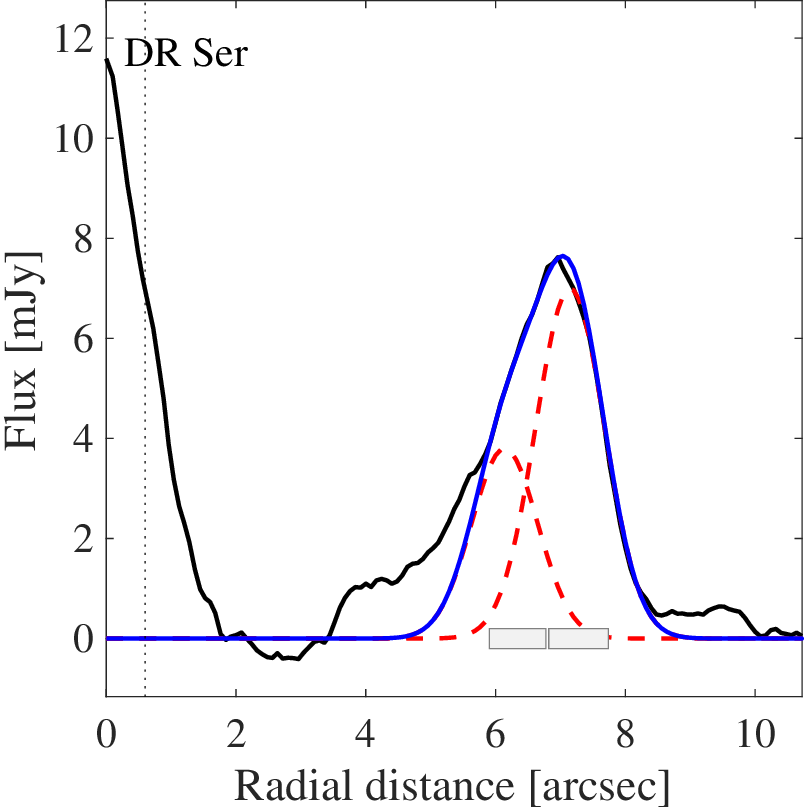} 
\includegraphics[width=4cm]{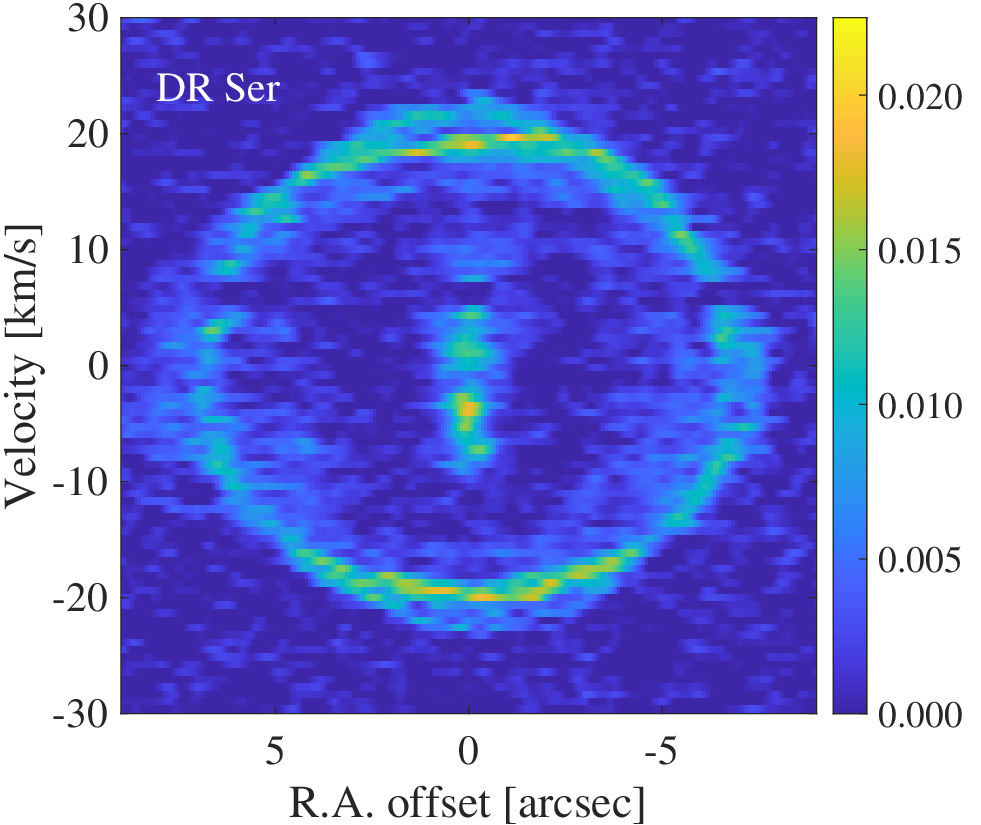}
\includegraphics[width=3.25cm]{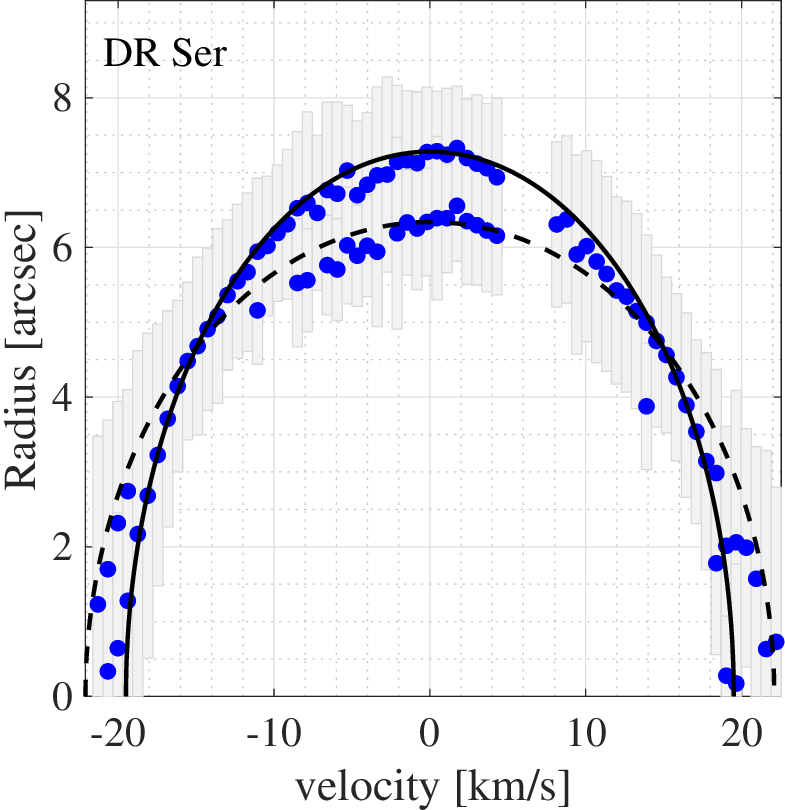} 
\caption{\emph{Left:} Moment 0 maps of the \COone~emission observed with ALMA (except U~Ant -- \COtwo). The color scale is given in Jy/beam. For each row the source is indicated in the top left corner. {The beamsize is indicated by the ellipse in the lower-left corner. }\emph{Middle-left:} AARPs of the shells at the \vlsr. The blue line shows the sum of two Gaussian shells (red dashed lines) with the parameters given in Table~\ref{t:shellresults}. The vertical dotted line indicates the radius of the aperture that was used to extract the ALMA spectrum {(corresponding to $\approx$10\% of the shell diameters)}. For U~Ant the negative flux indicates artefacts owing to insufficient feathering of the total-power data with the main-array observations. \emph{Middle-right:} Position-velocity diagrams in the R.A. direction at Dec-offset 0 (except V644~Sco at a position angle of 135\textdegree~to emphasise the structure outside the detached shell). The color scale is given in mJy/beam \emph{Right:} Radii of Gaussian fits to the azimuthally averaged radial profiles vs. velocity. The grey-shaded area shows the FWHM of the Gaussian fit. The black and dashed lines show the expected radius vs. velocity for the shells as determined in the AARPs and the spectra.}
\label{f:momentpv}
\end{figure*}

Figure~\ref{f:momentpv} shows the moment 0 maps, azimuthally-averaged radial profiles (AARPs) of the brightness distribution at the \vlsr, and position-velocity diagrams (PV-diagrams) of the \COone~emission observed with ALMA towards all sources. {For U~Ant we use the \COtwo~data for the analysis, which provides the highest angular resolution in the data presented by~\cite{kerschbaumetal2017}. Owing to the relatively high excitation temperatures estimated for the shells, the excitation will not lead to different information in the \COone~and\COtwo lines regarding the structure of the shells}. Spectra extracted centred on the star are shown in Figs.~\ref{f:spectrauant} to~\ref{f:spectradrser}. {The spectra are extracted from apertures that correspond to $\approx$10\% of the shell diameters (chosen to be large enough to contain all the present-day mass-loss but small enough to make the distinct two-peaked profile visible).} The full channel maps are shown in Appendix~\ref{a:channelmaps}. 

The detached shells are apparent in all sources, showing an overall spherical symmetry, and are consistent in size with previous observations~\citep[e.g.,][]{schoieretal2005}. In addition to the main detached shells, additional sub-shells are apparent for all sources. The spectra also show a splitting in velocity of the shells, indicating a faster shell at lower intensity, and a slower but brighter shell. This was already observed in U~Ant, and was interpreted as a result of the wind-wind interaction as the detached shell expands after having been created during the high mass-loss and expansion-velocity phase during a thermal pulse~\citep{kerschbaumetal2017}. 

While~\cite{kerschbaumetal2017} concluded that the detached shell around U~Ant splits into an outer, faster shell followed by a more filamentary and slower structure inside the shell, the new ALMA data motivates a re-interpretation of the kinematical and spatial structures of the shells. We therefore include U~Ant in the analysis of the shells here, using the same \COtwo\/ data as was presented in~\cite{kerschbaumetal2017}.

\subsubsection{Observed double-shell structures in U~Ant, V644~Sco, S~Sct, and DR~Ser}
\label{doubleshells}

\begin{figure*}[t]
\centering
\includegraphics[width=5.55cm]{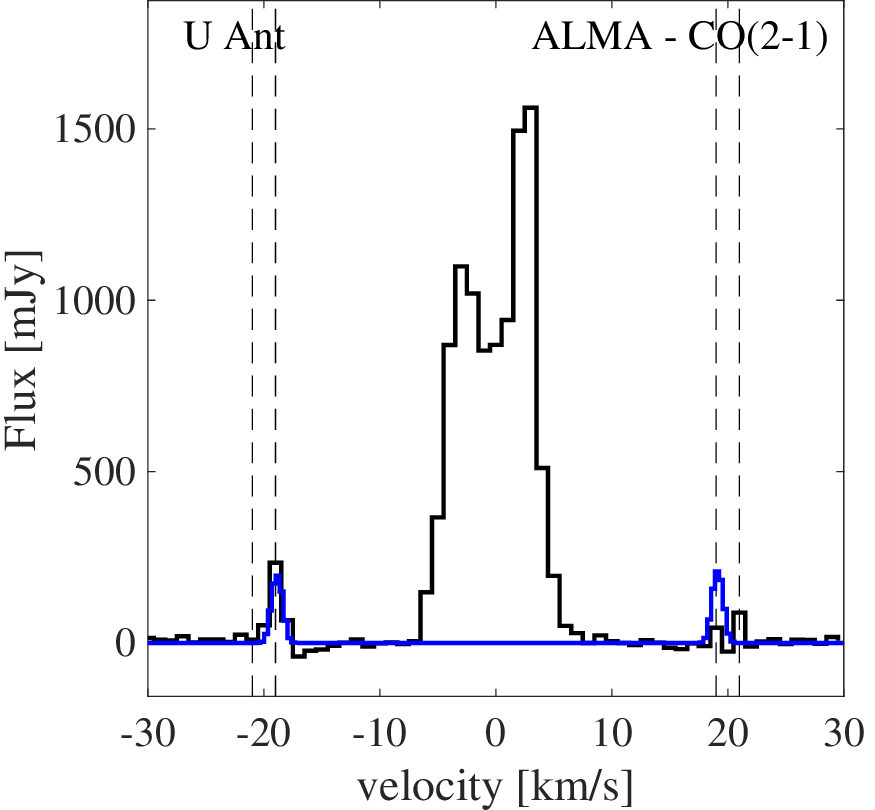}
\includegraphics[width=5.5cm]{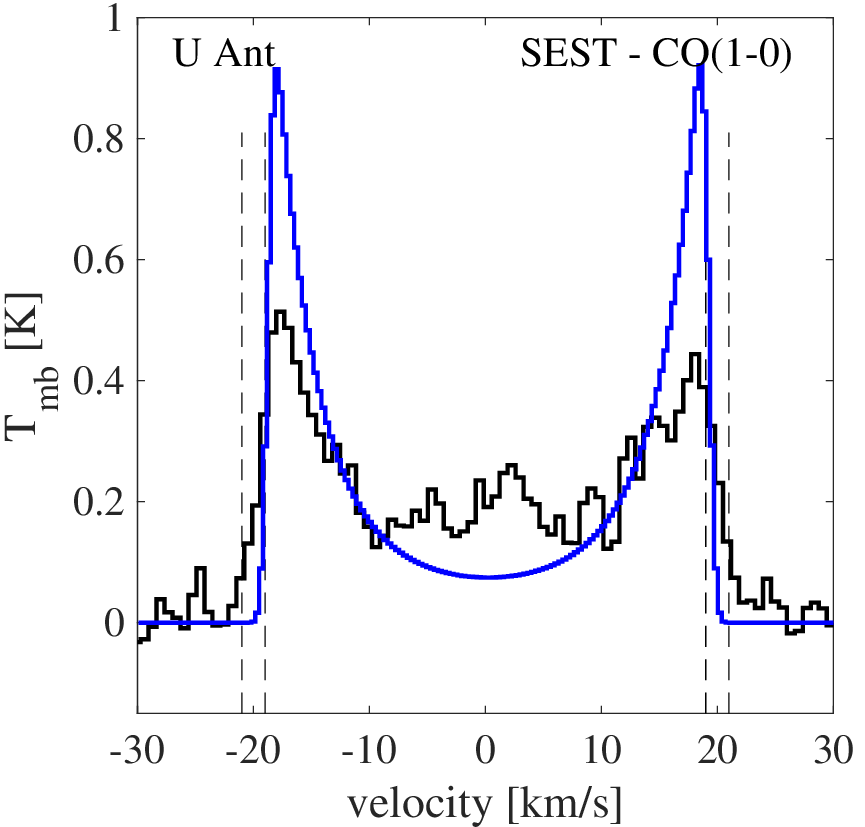}
\includegraphics[width=5.5cm]{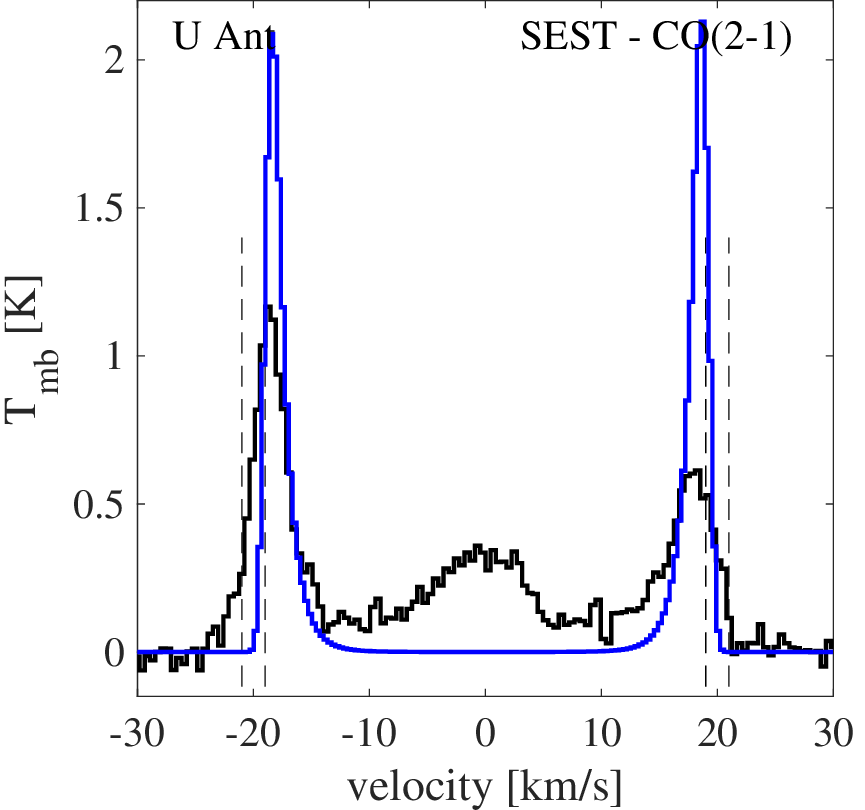} \\
\vspace{0.5cm}
\hspace{-11cm}\includegraphics[width=5.5cm]{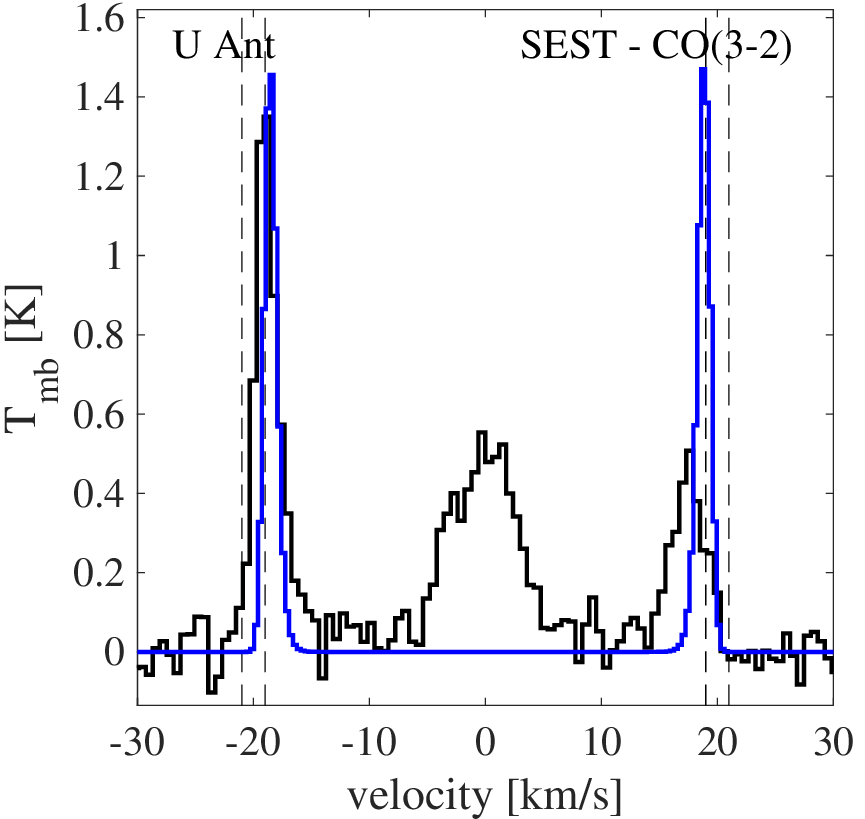}
\caption{ALMA and single-dish observations and the corresponding models (blue lines) for U~Ant, based on the values presented in Table~\ref{t:shellresults}. The top left panel shows the spectrum of the \COtwo~emission extracted from the ALMA data centred on the star, extracted from an aperture corresponding to $\approx$10\% of the shell diameter. The telescope and CO transitions for the single-dish data are indicated in the respective panels. The shell expansion velocities are indicated with vertical dashed lines and are used as input for the shells in Fig.~\ref{f:momentpv}, right column.}
\label{f:spectrauant}
\end{figure*}

\begin{figure*}[t]
\centering
\includegraphics[width=5.4cm]{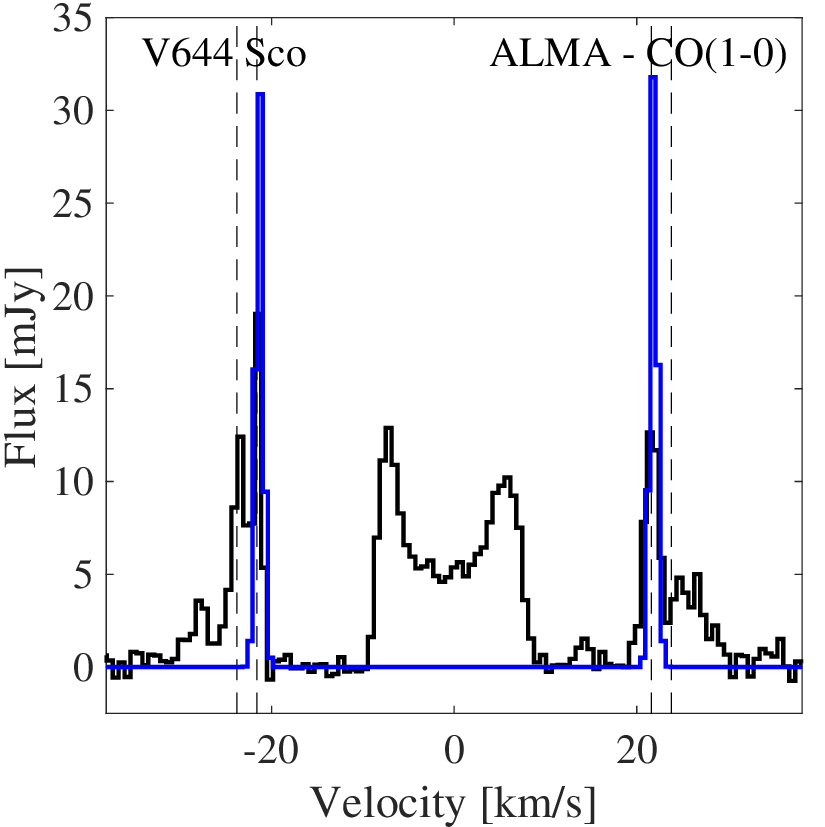}
\includegraphics[width=5.5cm]{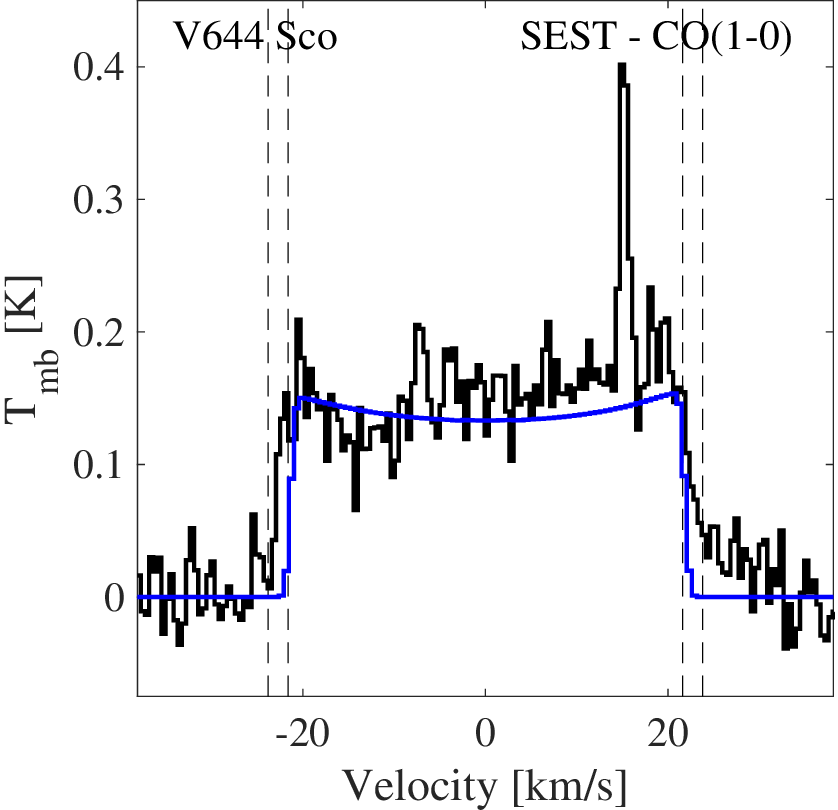}
\includegraphics[width=5.5cm]{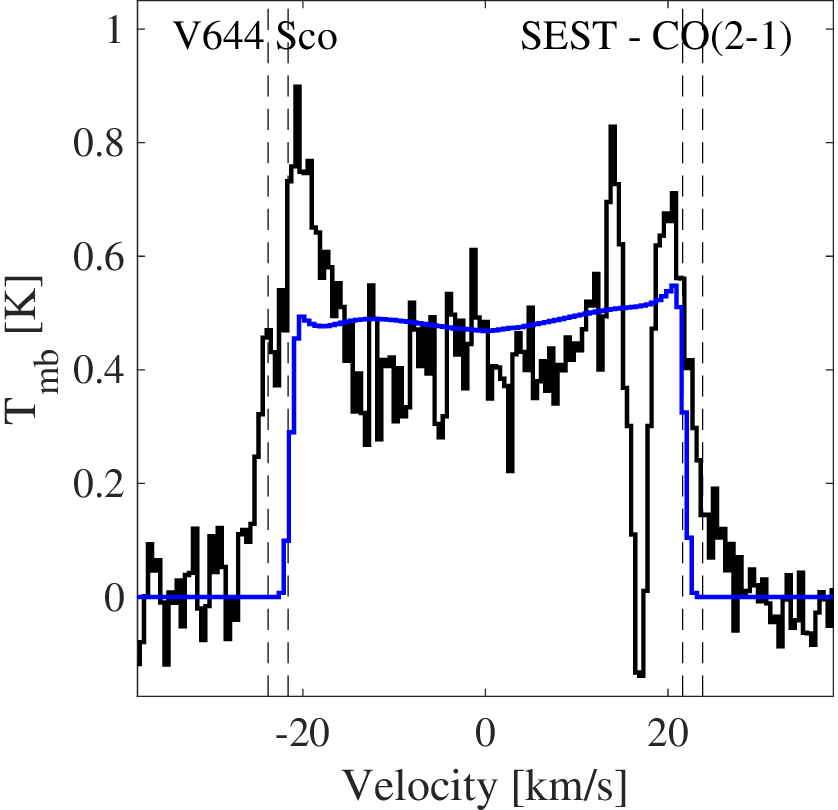}\\
\vspace{0.5cm}
\hspace{-5.5cm}\includegraphics[width=5.5cm]{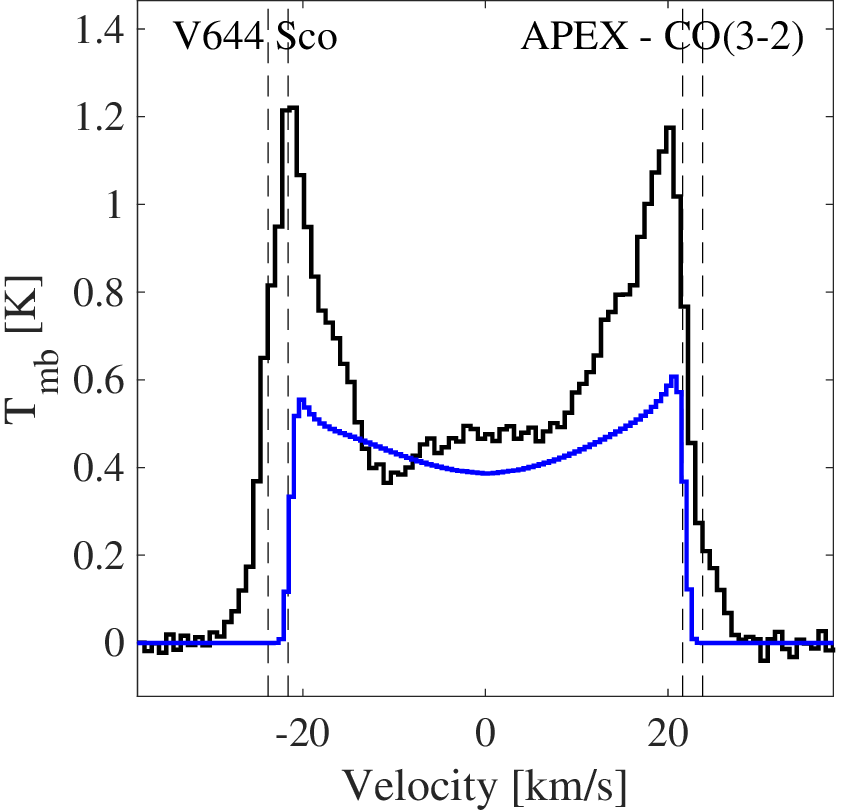}
\includegraphics[width=5.5cm]{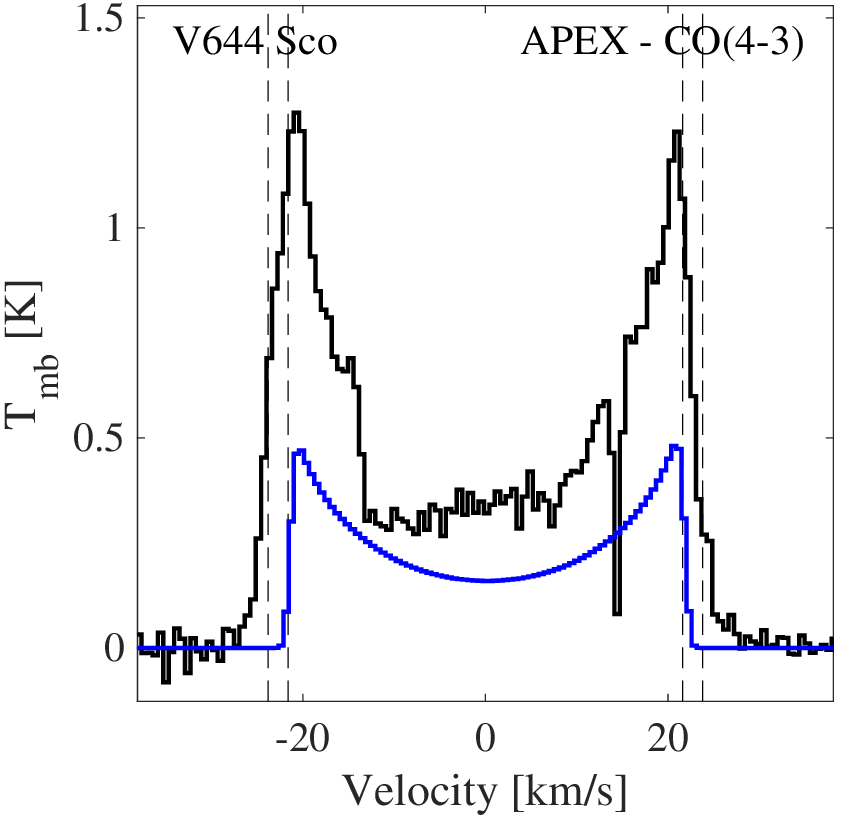}
\caption{Same as Fig.~\ref{f:spectrauant}, but for V644~Sco.}
\label{f:spectrav644}
\end{figure*}

\begin{figure*}[t]
\centering
\includegraphics[width=5.5cm]{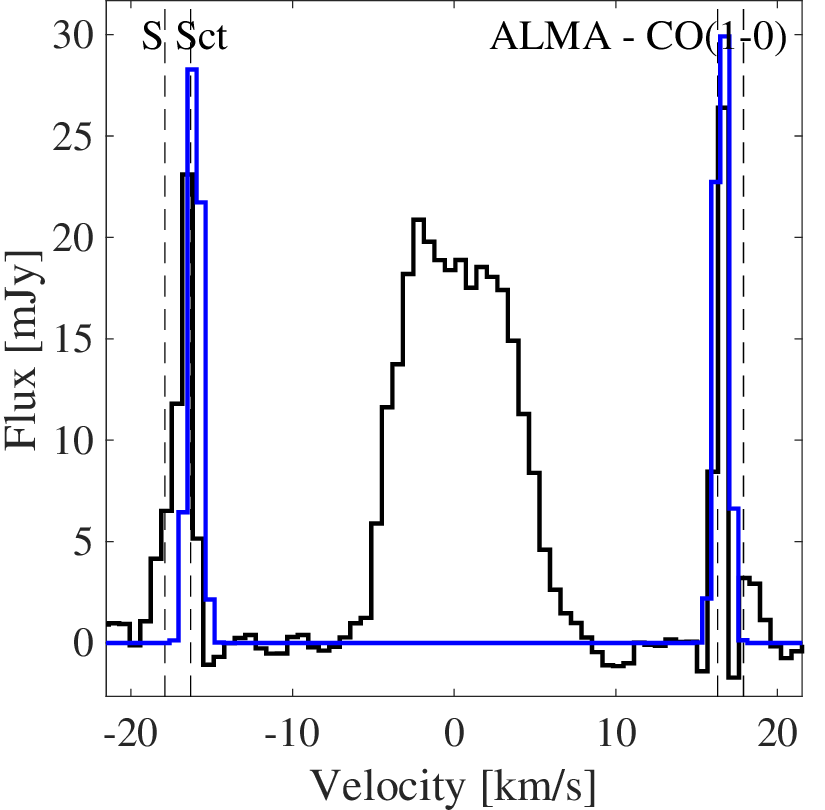}
\includegraphics[width=5.5cm]{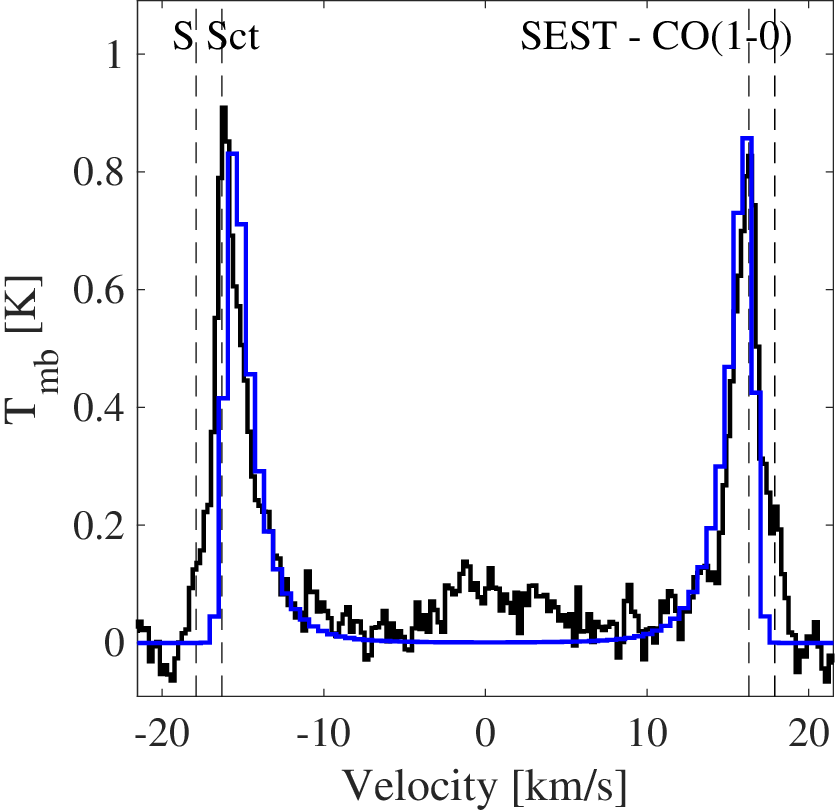}
\includegraphics[width=5.5cm]{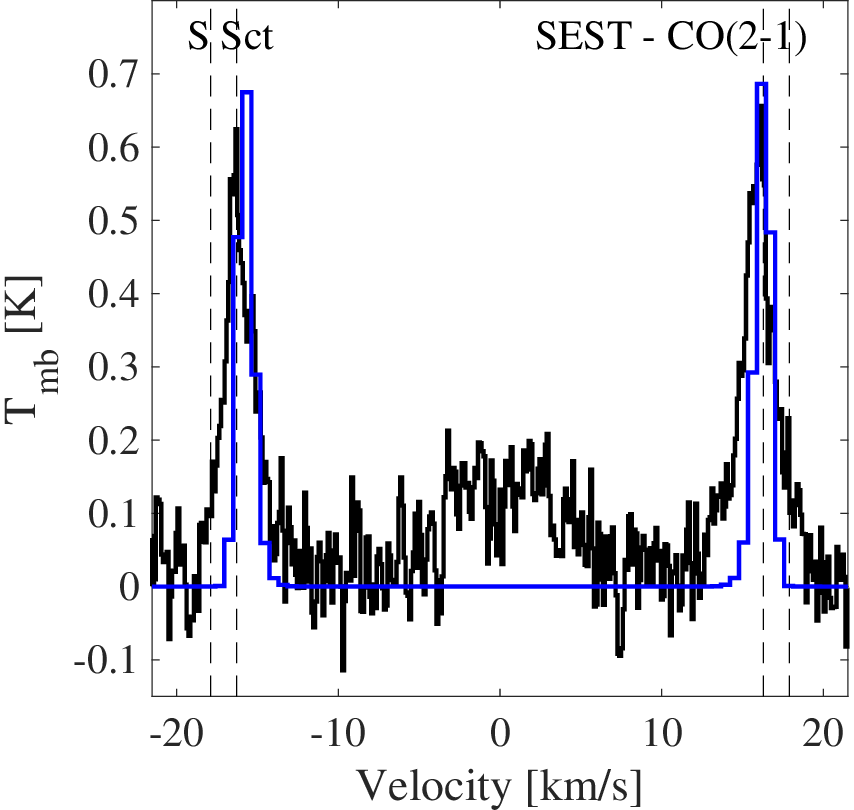}\\
\vspace{0.5cm}
\includegraphics[width=5.5cm]{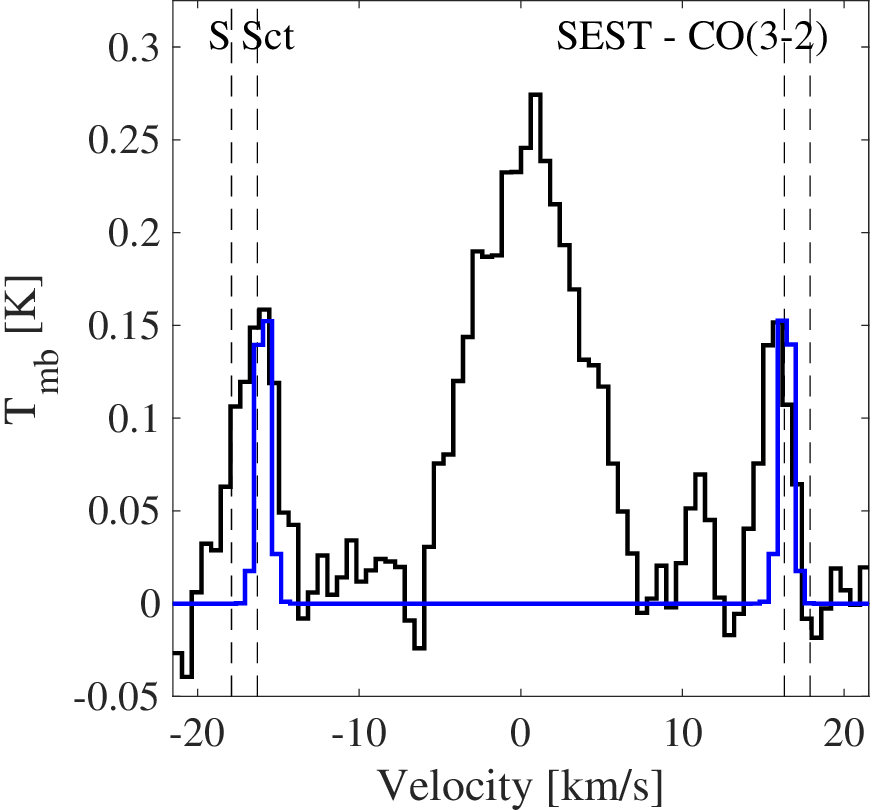} 
\includegraphics[width=5.5cm]{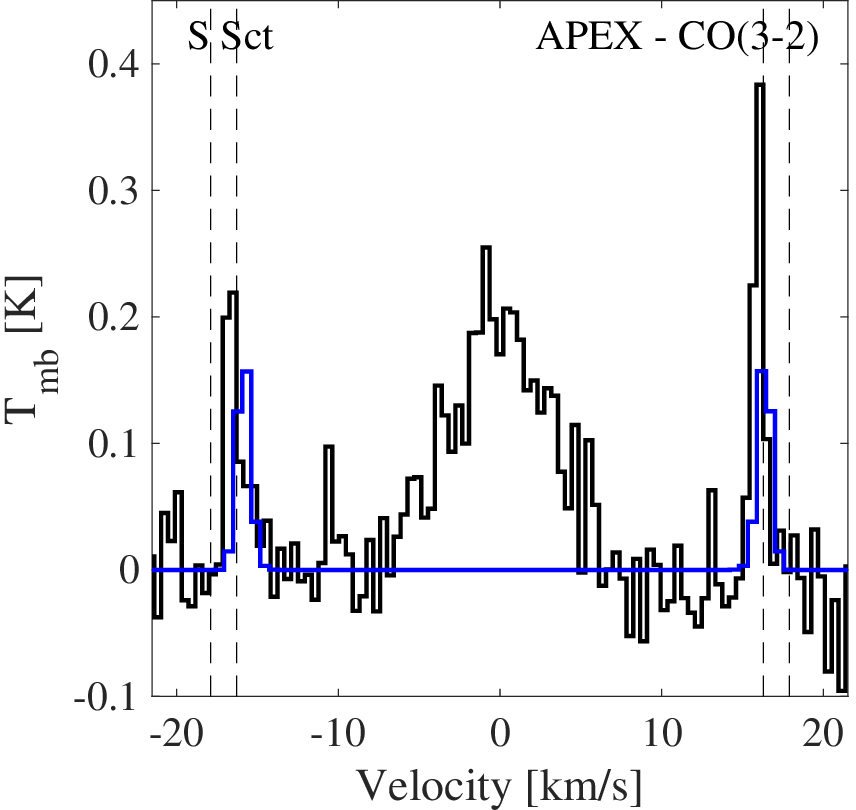}
\includegraphics[width=5.5cm]{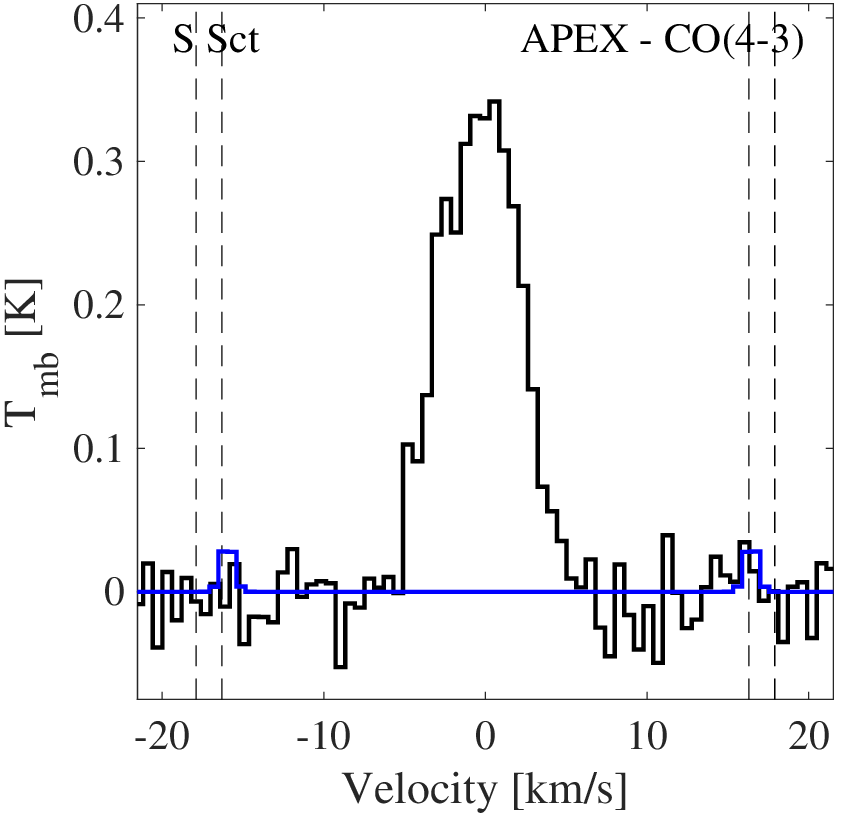}
\caption{Same as Fig.~\ref{f:spectrauant}, but for S~Sct.}
\label{f:spectravssct}
\end{figure*}

\begin{figure*}[t]
\centering
\includegraphics[width=5.3cm]{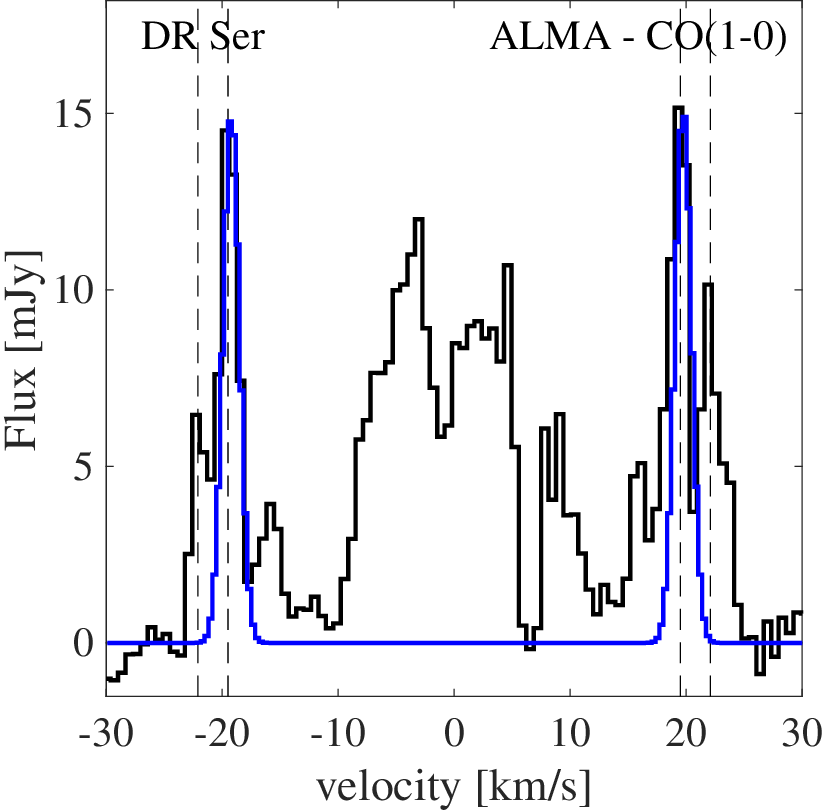}
\includegraphics[width=5.5cm]{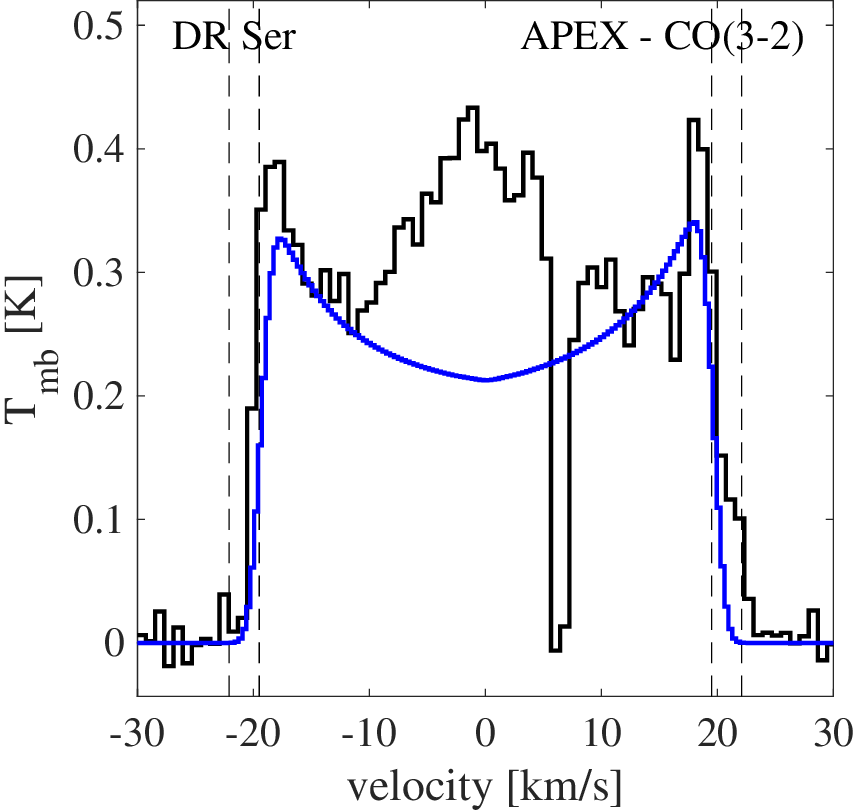}
\includegraphics[width=5.5cm]{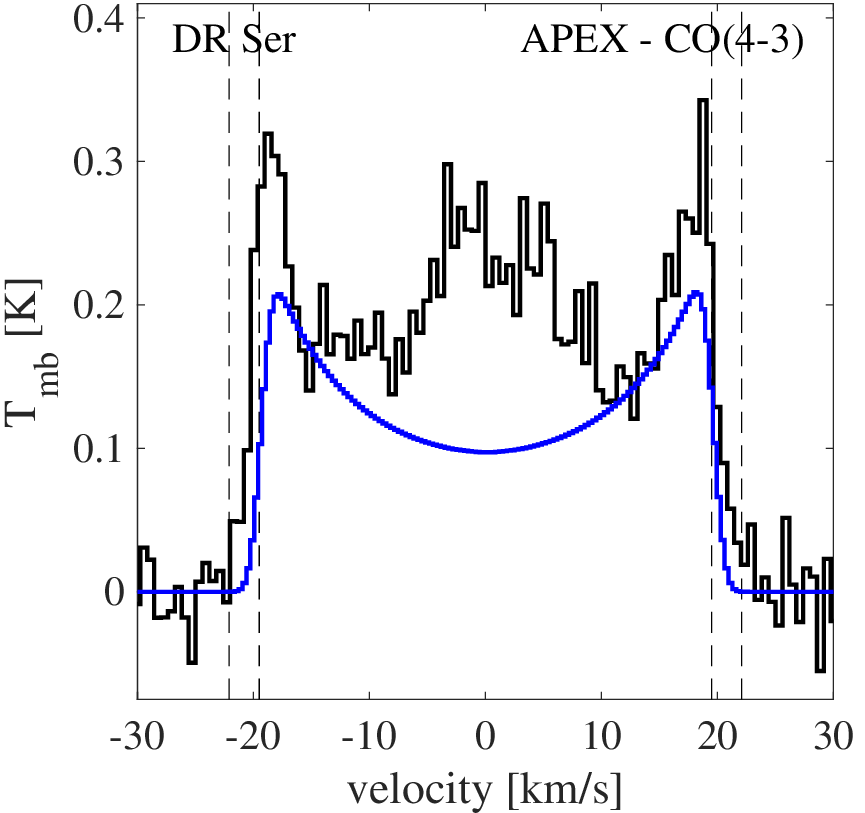}
\caption{Same as Fig.~\ref{f:spectrauant}, but for DR~Ser.}
\label{f:spectradrser}
\end{figure*}

As for U~Ant, double-shell structures are apparent in the emission around V644~Sco, S~Sct, and DR~Ser. All sources show a bright and nearly spherical outer detached shell, and a more filamentary, fainter shell structure just inside the outer shell. {Both the outer detached shell and the filamentary structures are centred on the star for all sources.} The spectra also show a splitting of the expansion velocity in two shells, roughly separated by 2\,\kms (as measured in the extracted ALMA spectra in Figs.~\ref{f:spectrauant} to ~\ref{f:spectradrser}; Table~\ref{t:shellresults}). The spectra imply that the faster component is fainter, corresponding to the filamentary structure of the inner shell, while the slower component is brighter, corresponding to the outer shell. 

To quantify the spatial and dynamical structures in the double-shells better, we create azimuthally averaged radial profiles (AARPs) in each velocity bin for all sources and fit two gaussian shells to the profiles, providing the projected shell radii as a function of velocity in radius vs. velocity plots (RV-plots; Fig.~\ref{f:momentpv}, right). The same method was used to determine the shell radius and velocity in the ALMA Cycle 0 data for R~Scl~\citep[albeit for only one gaussian shell;][]{maerckeretal2012}. In principle, this provides the same type of information as the PV-diagrams in Fig.~\ref{f:momentpv} (middle-right). However, while the PV-diagrams are only along a cut in one direction, the RV-plots use information in the entire data set at each velocity. 

The largest \emph{spatial} separation of the two observed shells in the RV-plots is expected to be at the systemic velocity (the stellar \vlsr), that is, where we see the emission in the plane of the sky. Hence, the fits to the AARPs at the stellar \vlsr~provide the radii ~\Rshell~and widths ~\dRshell~(the FWHM of the gaussian shell) of the shells. 

Note that, although the shells are generally close to spherically symmetric, deviations from symmetry artificially broaden the shell-profiles. This results in an average shell radius and a broader shell width when creating an AARP over the entire 360 degrees. An average radius suffices for the analysis of the shells we provide here, {and we therefore measure the shell radii~\Rshell~in the AARPs over 360 degrees. However, in order to determine values that more accurately represent the FWHM of the shells, we measure the shell widths~\dRshell~over position-angles covering only a single two-degree interval. The respective position angle ranges for each source are given in Table~\ref{t:shellresults}}. The measured FWHM of the shells are additionally deconvolved by the observational beams. Table~\ref{t:shellresults} shows the average radii of the shells and the deconvolved FWHM of the gaussian fits.

The largest \emph{velocity} separation is expected at the caps of the shells along the line of sight (that is, approaching a projected radius of 0). In the case of a faster inner shell and a slower outer shell, the projected radii would be separated at the \vlsr, and then merge into each other progressing through the velocity channels towards the shell-expansion velocities, to then separate again to reach their respective expansion velocities at the radius = 0. 

This behaviour can be seen for U~Ant, S~Sct, and DR~Ser, where the measurements are initially separated, then merge, to finally separate into two shells again at the extreme velocities. For the shells around V644~Sco the situation is less clear, as the inner shell is almost as large as the outer shell. However, also here the data is consistent with a faster inner and slower outer shell. 

In Fig.~\ref{f:momentpv} (right), we indicate the location of theoretical, spherical shells consistent with the parameters of the shells presented in Table~\ref{t:shellresults} (where Radius=\Rshell$\times \sqrt{1-({v}/\varv_{\rm{sh}})^2}$).  For the shells with the largest angular sizes, U~Ant and S~Sct, the two shells are clearly resolved. For the smaller shells, V644~Sco and DR~Ser, the separation is less apparent. However, for all sources the measured radii and expansion velocities are consistent with two nearly spherical shells, with a faster inner and a slower outer shell.

Given the radii and expansion velocities of the shells, we can estimate the age of the shells using their velocities as upper and lower limits. Since the outer emission likely traces material that has slowed down since it was ejected, the velocities from the outer shells provide upper limits to the shell ages. The inner emission has not been as affected by the interaction with the previous winds, and the velocities from these shells result in upper limits to the shell ages. The derived age ranges are indicated in Table~\ref{t:shellresults} and span ages from $\approx$1000 years after a thermal pulse up to $\approx$8000 years. The observations hence potentially probe a significant fraction of the pulse and inter-pulse development.

\subsubsection{Arc-structures around R Scl}
\label{rsclarcs}

R~Scl was one of the first sources to be observed with ALMA in Cycle 0. The observations with the early science array showed the detached shell, as well as an unexpected spiral structure extending from the shell inwards towards the star~\citep{maerckeretal2012,maerckeretal2016a}. The interpretation of the observed structure was that the wind around R~Scl is shaped into a spiral by a companion star. 
The Cycle 0 data was, however, strongly limited by spatial resolution and sensitivity ~\citep[only 16 antennas in a compact configuration with resolutions of $\approx$3\arcsec~and $\approx$1\farcs3 in \COone~ and \COthree, respectively, and without the Atacama Compact Array (ACA) or total power observations;][]{maerckeretal2016a}. 

The new data presented here with higher spatial resolution confirms the structures in the volume between the detached shell and the star. However, while the data is consistent with the observations in Cycle 0, the improved sensitivity shows that the structures around R~Scl rather are a series of intersecting arcs. 

While these structure may still be created by a (binary) companion~\citep[e.g.,][]{cernicharoetal2015,decinetal2020}, the derivation of binary parameters and the mass-loss history is substantially more complicated than reported by \cite{maerckeretal2012}. The analysis of the entire Cycle 0 dataset, including radiative transfer models of single-dish observations, however, confirmed the conclusion that the mass-loss rate from R~Scl must have remained high after the thermal-pulse~\citep{maerckeretal2016a}.

With the new data it is possible to make a more detailed analysis of the detached shell around R~Scl and the arc-structures inside the shell. For R~Scl, the shell has an expansion velocity of $\approx$14.5\,\kms, consistent with previous estimates. The arc structures can also be seen in the PV-diagrams and RV-plots, from just inside the shell to distances of only a few arc-seconds from the star. The arcs are likely not complete, concentric, and spherical shells, but rather partial shells. The ellipses they form in the PV-diagram may therefore be incomplete and deviate from the ideal shape of an ellipse. 

The PV-diagram and RV-plot show that the inner arcs converge on higher expansion velocities ($\approx16$\,\kms) than the detached shell. Only the innermost arcs ($\lesssim$ 4\arcsec) have velocities that are slower than the detached shell.

\subsection{Additional structures and features}
\label{s:other}

\subsubsection{V644~Sco: additional shells}
\label{s:v644other}

In addition to the observed double-shell structure, V644~Sco shows emission from clear shell-like structures at distances from the star outside the detached shell. There appears to be a dominating shell with a radius of $\approx$12\farcs5 expanding at a velocity of $\approx$25\,\kms~(Figs.~\ref{f:momentpv} and ~\ref{f:v644maps}). Additional substructures in the form of arcs are also present. 

While the detached shell is a complete, spherical shell, this outer shell around V644~Sco is only observed towards the north-east, extending to the line-of-sight (the outer shell can be seen in the extracted ALMA spectrum centred on the star as high-velocity peaks). Note that also the detached shell is significantly brighter in the same direction as this outer shell. 

It is not clear what the origin of this shell is. Assuming a velocity of 25\,\kms, the time between this outer arc and the detached shell corresponds to $\approx$500 years. This essentially excludes the possibility that the arc would have been created in an earlier thermal pulse. Variation of the mass-loss rate before the thermal pulse also seems unlikely, since one would expect this to be symmetric around the star. Finally, interaction with the ISM seems unlikely, as we would expect that to create one interaction region, and not multiple shells, and the shape of the wind-ISM bow-shock generally flares out more~\citep{maerckeretal2022}. The nature of these outer arcs will have to be investigated further in the future.

\subsubsection{DR~Ser: a filled shell}
\label{s:drserother}

\begin{figure*}[t]
\centering
\includegraphics[width=5.5cm]{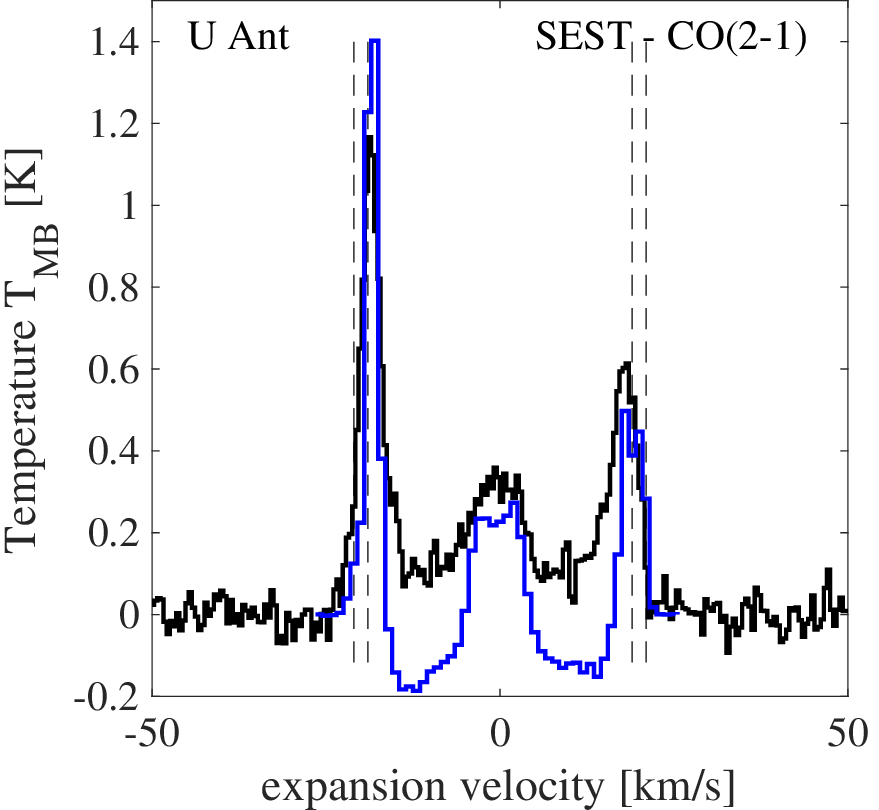} 
\includegraphics[width=5.3cm]{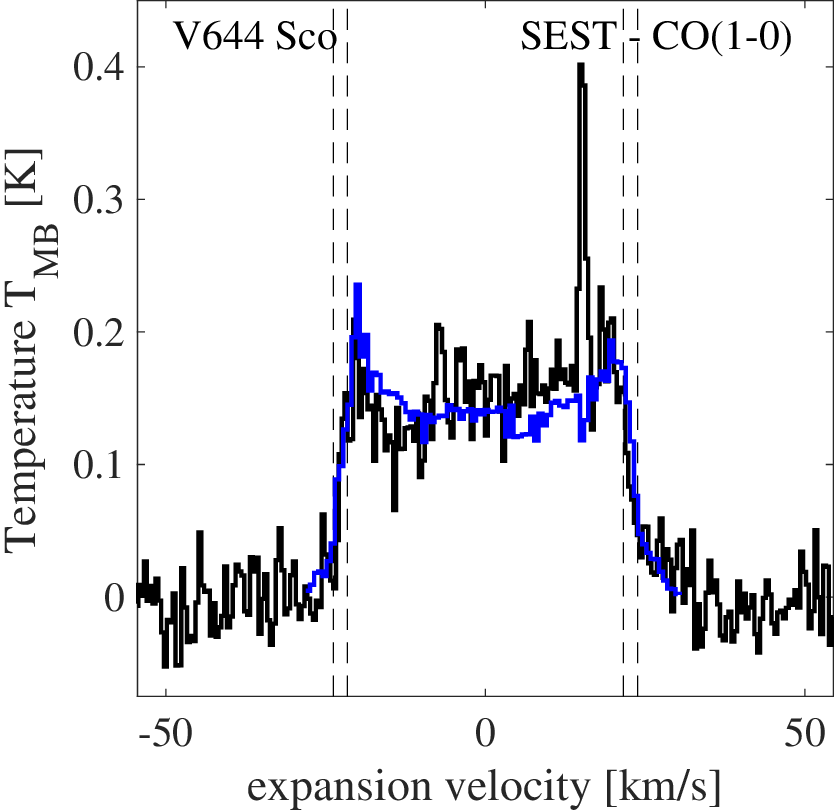}
\includegraphics[width=5.3cm]{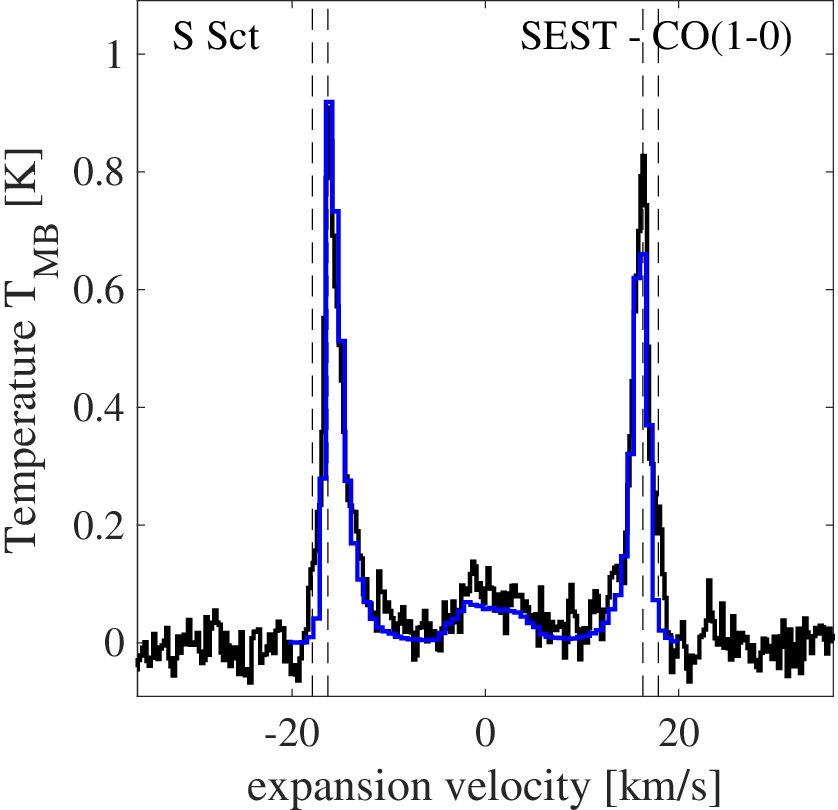}
\caption{ALMA observations on \COtwo~(U~Ant) and \COone~(V644~Sco and S~Sct) convolved with the respective SEST beams (blue) compared to the SEST spectra (black).}
\label{f:sdcompare}
\end{figure*}

The AARPs for the shells generally show no emission inside the shell (that is, the AARPs go to 0 flux). {{To address the possibility of resolved-out flux, we compare the ALMA data convolved with the single-dish beam to the single-dish spectra (Fig.~\ref{f:sdcompare}). Unfortunately we do not have spectra in \COone~towards DR~Ser. However, for V644~Sco and S~Sct the comparison shows that ALMA is not resolving out any flux, while for U~Ant the opposite is clearly the case~\citep[caused by insufficient combination of the data to fill in the zero-spacing; see][]{kerschbaumetal2017}. Aside from the data combination issues for U~Ant, the ALMA observations hence show emission on all scales.}}

\begin{figure*}[t]
\centering
\includegraphics[width=18cm]{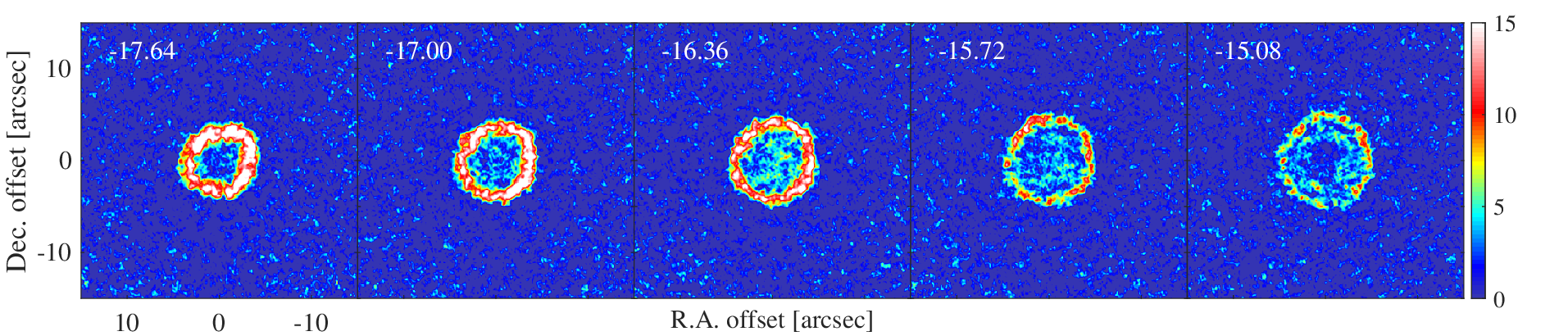} \\
\caption{Channel maps of DR~Ser in \COone. The colorbar is given in mJy/beam. The panels show velocity bins between -17.64\,\kms~and -15.08\,\kms, showing channels in which the shell around DR~Ser seems at least partially filled. T The full range of channel maps is shown in Fig.~\ref{f:drsermaps}. }
\label{f:drsermapsrange}
\end{figure*}

\begin{figure}[t]
\centering
\includegraphics[width=6cm]{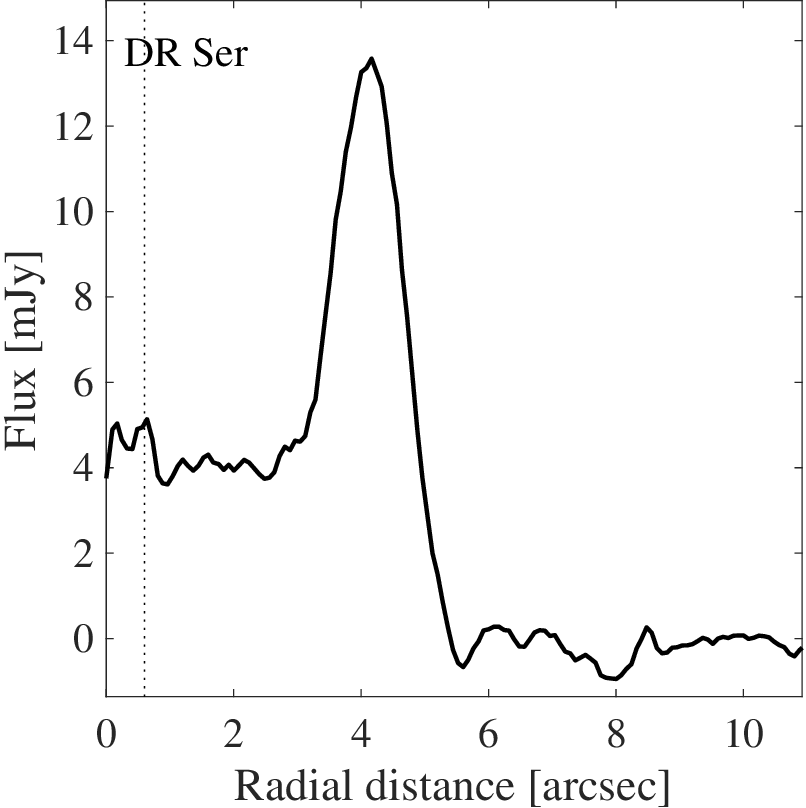} 
\caption{The AARP for DR~Ser at -16.36\,\kms. The vertical dotted line indicates the radius of the aperture that was used to extract the ALMA spectrum. }
\label{f:drserfilled}
\end{figure}

{{Comparing the emission towards V644~Sco and S~Sct to DR~Ser, we see that}} for DR~Ser there is clear smooth emission also at radii smaller than the shells. Significant emission inside the shell can be seen at all velocities, although it becomes clearer when approaching the shell expansion velocities {{(Figs.~\ref{f:drsermapsrange} and~\ref{f:drserfilled} show channel maps between -17.64\,\kms~and -15.08\,\kms~, and the AARP at -16.36\,\kms, respectively)}}. The maximum recoverable scale based on the array configuration (Table~\ref{t:almaobs}) in the observations of DR~Ser is $\approx$\,25\arcsec -- significantly larger than the shell diameter (15\arcsec). Any smooth emission on the scale of the detached shell should therefore not be resolved out{{, and this is confirmed for V644~Sco and S~Sct}}. However, the shell is not clearly filled-in at the \vlsr, {{and some minor negative features in the AARP of DR~Ser in Fig.~\ref{f:momentpv}}} indicate that possibly some of the flux is resolved out.

For R~Scl emission inside the shell is apparent in the form of the arcs possibly created by the shaping of the post-pulse wind by a binary companion. In the absence of a companion, the post-pulse wind is likely to be comparatively smooth. Although the emission inside the shell from DR~Ser is significantly weaker compared to that of R~Scl, it is possible that the emission seen inside the shell of DR~Ser also is a consequence of the post-pulse mass loss. This would indicate that the post-pulse mass loss remains significant also in this source. 

DR~Ser has the youngest shell in our sample. It is hence possible that the shell in this case has not yet completely detached. At the same time, V644~Sco is only somewhat older (and younger than R~Scl) and does not display emission inside the shell. 

\subsection{Radiative transfer models}
\label{s:radmodels}
Radiative transfer models allow us to derive the masses and temperatures in the detached shells. This was done previously using single-dish observations of CO~\citep{schoieretal2005}, providing an at that time comprehensive view of the properties of detached shells around carbon stars. 

However, the models were affected by a limited number of observed CO transitions and a lack of spatial resolution. For R~Scl, the modelling was repeated using the ALMA Cycle 0 data to spatially constrain the model, combined with single-dish transitions (that do not spatially resolve the shell width) to constrain the shell mass and temperature~\citep{maerckeretal2016a}. Also~\cite{kerschbaumetal2017} used  spatially resolved ALMA observations of U~Ant in combination with single-dish observations to constrain the properties of the shells. 

In a similar way, we use the spatial and velocity constraints obtained from the ALMA data here to model the emission from the shells. 

Initially, we attempted to model the emission from two shells. However, it is likely that the emission does not trace two shells, but rather inner and outer regions of one shell (see Sect.~\ref{s:hydrocompare}), making the connection between H$_{\rm{2}}$ density and CO emission uncertain. In addition, the complexity of the velocity fields and the evolution of the expansion velocities complicates the modelling. Hence, we focus on only the outer shell, and determine whether the emission modelled in \cite{schoieretal2005} is consistent with the brighter, outer shells observed in the ALMA data.

We use a 1-dimensional radiative transfer code based on the Monte-Carlo technique~\citep{schoierco2001}. The same code was used in \cite{schoieretal2005} and \cite{kerschbaumetal2017}. For each source, the input to the models is the radial density distribution consisting of a step-function centred on the shell radius~\Rshell, with the shell width~\dRshell ~and (constant) expansion velocity \vshell. The values for \Rshell, \dRshell, and \vshell~are taken from Table~\ref{t:shellresults} for the outer shells. The temperature in the shell~\Tshell~is assumed to be constant. Masses and temperatures for the shells are taken from \cite{schoieretal2005} and \cite{kerschbaumetal2017} and are presented in Table~\ref{t:shellresults}. The models assume a constant fractional abundance of CO relative to H$_2$ of \fCO=$10^{-3}$ and a turbulent velocity of 0.5\,\kms. Values for the stellar luminosity, effective stellar temperature, and distances are taken from~\cite{schoieretal2005}.

 {The observations indicate a complicated evolution of the mass-loss rate and expansion velocity throughout the thermal pulse cycle. Since this paper focuses on the kinematical and spatial structures of the detached shells, adding the present day mass-loss would add to the complexity, without adding much more information than provided in~\cite{schoieretal2005}}. We therefore do not include a present-day mass-loss. The models are compared to the ALMA spectra extracted from the position centred on the star, and the single-dish observations in Table~\ref{t:sdobs}. 

Figures~\ref{f:spectrauant} to~\ref{f:spectradrser} show the resulting models. Overall the models fit reasonably well. Also the added higher-$J$ transitions of CO are consistent with the modelled shell masses and temperatures. This confirms that the emission in the spectra is dominated by the outer shell. For U~Ant we get a better fit to the data if we use a mass of $1\times10^{-3}$\,\Msun, and for V644~Sco the ratio between the higher-$J$ and low-$J$ transitions indicates that the temperature in the shell might be higher than 170\,K. 

At the same time, clear deviations in the line profiles from what is expected from a simple shell-structure can be seen, in particular for V644~Sco and DR~Ser. For these sources the observed \COthree~and \COfour~lines show sharp double-peaked profiles with a mostly flat-topped emission profile in between the peaks. While a double-peaked profile is expected considering the smaller beam compared to the shell radii, the line profile should have a much rounder shape in between peaks, albeit affected by the presence of a present-day mass-loss.

It is clear that the observed structures are more complicated than simply two shells expanding at separate velocities. Additional structures outside the shells (as is seen in V644~Sco), possible pre-pulse mass-loss, significant post-pulse mass-loss (as is possibly seen in DR~Ser), a present-day mass-loss, and additional structures (for example the binary-induced arcs in R~Scl), all with unknown velocity profiles, contribute to the observed spectra, and complicate the interpretation.

In addition, it is unlikely that the CO emission is tracing the (entire) density structure of the detached shells (see Sect.~\ref{s:discussion} for a discussion on and interpretation of the observations). While our models show that the masses determined in earlier work are consistent with the emission from the outer shell, this emission likely misses a significant amount of mass in the shells. Hence, any mass estimate from the radiative transfer models is only a lower limit and very uncertain.

\section{Discussion: the evolution of detached shells}
\label{s:discussion}

\subsection{Comparison to hydrodynamical models: single vs. double shells}
\label{s:hydrocompare}
For the sources discussed here, we find that previous models of the masses in the detached shells are consistent with CO emission from the thin outer shells. The masses are also consistent with the mass loss predicted during a thermal pulse~\citep[e.g.][]{steffenco2000,schoieretal2005,mattssonetal2007,karakas2010,maerckeretal2016a,kerschbaumetal2017}

However, the ALMA observations consistently show that the CO emission from the detached shells splits into two shells, both spatially and dynamically, and the inner shells appear to be moving faster than the outer shells. The hydrodynamical models by~\cite{steffenco2000} and~\cite{mattssonetal2007} show a velocity gradient across the shell, with a lower velocity at larger radii, and a higher velocity at smaller radii. The difference in velocities across this gradient is consistent with the difference between the inner and outer shells measured here ($\approx$ 1--3\,\kms). 

Further, early after the formation of the shells, the models predict a larger gradient in velocity across the shell that gradually declines as the inner part of the shell is slowed-down by internal pressure. At the same time, the shells broaden and maintain a roughly constant \dRshell/\Rshell. In the sources we observe in this paper, the separation in velocity indeed declines as the shells get older, while the spatial separations compared to the shell radii remain constant (Table~\ref{t:shellresults}). 

On the other hand, the hydrodynamical models do not predict double-shell structures in H$_2$~\citep{steffenco2000,mattssonetal2007}. Instead, the theoretical shells are single shells either similar to a Gaussian density distribution, or with a more flat-topped density distribution. Since the velocity differences and separations between the CO emitting shells are consistent with the velocity gradient and widths of the H$_2$ shells in the hydrodynamical models, it seems likely that the CO emission is tracing the front (outer shell) and back (inner shell) parts of the detached shells, with a, possibly, significant amount of mass in between the CO-shells. The separation measured in the observed shells would in that case constrain the width of the detached shells, instead of describing two separate shells.

It is not immediately obvious how to reconcile the single-shell (H$_2$) structures in the hydrodynamical models with the observed double-shell (CO) structures. As mentioned previously (Sect.~\ref{s:results}), the observations only trace the distribution of CO {emission}, while the hydrodynamical models model the density distribution of \Htwo. If indeed the CO observations are only tracing the inner and outer edges of the detached shells, with an unknown \Htwo~distribution in between, { the centre of the shell must be essentially devoid of CO (the optical depths are $\lesssim$\,1.3 and hence too low for the structures to be explained by high optical-depth effects)}. {At the distances of the detached shells, CO is not expected to be photodissociated by the interstellar radiation field~\citep{saberietal2019}. A possible explanation for the distribution of CO could be the destruction of CO as the gas crosses the shock, after which it reforms in the post-shocked gas.} Dissociation of CO needs temperatures in excess of 3000\,K. While~\cite{mattssonetal2007} obtain temperatures of "a few thousand K" in the contact discontinuity of the shock in the shell, these temperatures depend critically on the assumed and very uncertain cooling and heating processes.

We investigated the possibility to model a single-shell with a \fCO$(r)$ that reduces the amount of CO in the centre of the shell and a velocity gradient across the shell. However, since the degree, width and position of the reduction of CO are unknown, as well as the shape of the single-shell, the problem contains too many free parameters and is too degenerate to model to obtain meaningful results. Any additional emission inside and outside the shells further complicates the problem (see Sect.~\ref{s:radmodels}). It is clear though that in the case of a single shell, the shell-masses derived through CO observations are lower limits to the total shell masses. 

For a single \emph{Gaussian} shell of H$_2$, the CO emission would originate from the wings of the Gaussian distribution, and the total mass may be significantly higher. Considering that the derived masses already now are consistent with what is expected of the mass-loss rate during a thermal pulse, such a large total mass is unlikely, even if mass is swept-up from the pre-pulse mass-loss. A more flat-topped \Htwo~density distribution is more likely, similar to what is shown in, e.g., Fig.~10 in~\cite{mattssonetal2007}.

\cite{olofssonetal2015} observed CI towards R~Scl with APEX. Although the authors conclude that the observed CI emission is indeed dominated by emission from the shell, the results are consistent with dissociation of all carbon-bearing molecules other than CO. However, the models only consider one Gaussian shell with a constant CO abundance. In light of the results reported in this paper, it may be worth revisiting the results by~\cite{olofssonetal2015}.

\subsection{Constraints on the thermal pulse cycle}
\label{s:tpconstraints}

During and after a thermal pulse, the mass-loss rate and expansion velocity of the wind are expected to first increase for a few hundred years, to then both decrease significantly, creating a shell of dust and gas that expands away from the star and detaches. 

In the hydrodynamical models by \citet{steffenco2000,mattssonetal2007}, the mass-loss rates decline to their pre-pulse values within $\approx$1400 years, while the expansion velocity reaches its pre-pulse value within $\approx$3800 years~\citep[for the detailed evolution, see Figs. 2 and 3 in][]{mattssonetal2007}. 

In all modelled cases, the shell structure remains stable over a period of ~$>10^4$ years. The age ranges determined for the shells in our study span all of these scales, from the youngest sources possibly still affected by relatively high post-pulse mass-loss-rates and expansion velocities, to the older sources that have completely detached shells.

It is interesting to note that for R~Scl the multiple shells inside the detached shell indicate that the expansion velocity remains high for a significant time after the pulse,  declining only recently ($\approx$400 years ago). R~Scl has a detached shell with intermediate age in our sample ($\approx$1500 years), implying that the velocity and mass-loss rate after the pulse remains high for this time. This is not in disagreement with the theoretical models, albeit at the upper end of what they predict. 

For DR~Ser, the youngest source in the sample, there are signs that the shell may still be filled, implying that also here the mass-loss rate remains reasonably high until $\approx$1000 years after the shell was created. As mentioned in Sect~\ref{s:drserother} though, the shell around V644~Sco has an age that lies between those of the shells around DR~Ser and R~Scl, but here there are no signs of a high post-pulse mass-loss rate or expansion velocity.

Although some basic aspects of the evolution of the shells can be constrained with these observations, it is clear that a straightforward interpretation of the implications on the thermal-pulse cycle is not possible, requiring more detailed modelling.

\subsection{The need for comprehensive modelling}
\label{s:newmodels}
The spatial and dynamical structures of the shells are a result of the conditions during the thermal pulse and the evolution of the shells. The observed double-shell structures and velocity distributions show that it is not straightforward to derive parameters such as mass-loss rate or expansion velocity during the thermal pulse, or their subsequent evolution. 

Ideally one would comprehensively model the complete (observed) double-shell structure, including possible pre-pulse mass-loss and post-pulse mass-loss, with varying velocities. This would require a better theoretical understanding of the structures in the shells from hydrodynamical models, as well as additional observations to constrain both the abundance distribution of CO and the distribution of \Htwo. 

The models by \cite{steffenco2000} and \cite{mattssonetal2007} are only in one dimension. Full 3-dimensional models, coupled with radiative transfer including dust and the chemistry of CO, are necessary to properly describe the cooling and heating in the shells, as well as instabilities and the filamentary structure of the inner shell~\citep{mattssonetal2007}. 

Such a comprehensive modelling would constrain the evolution of the shells, and determine the conditions for creating the shells during the thermal pulse.

\section{Conclusions}
\label{s:conclusions}

In this paper we present ALMA observations in \COone~towards five carbon-rich AGB stars with detached-shells of gas and dust. We determine the dynamical and spatial properties of the observed shells, and investigate how these relate to previous observations with single-dish telescopes. 

While the masses and the spatial/kinematic properties are consistent with previous investigations, we find a splitting of the shells both in velocity and space in the emission of CO. Similar splitting was observed for U~Ant previously. With the new data it appears that a stable, double-shell structure in CO emission is common for most, if not all, detached-shell sources of this type, and that the determined masses from previous work trace the brighter, slower, outer shells. Assuming that the shells are formed during the period of high mass-loss and high expansion velocity during a thermal pulse of the AGB star, this structure must be maintained through the wind-wind interaction with the pre-pulse wind.

However, the observed double-shell structure likely does not trace the density of \Htwo, but rather the distribution of CO in the inner and outer edges of the shells. This requires a significant destruction of CO in the centre of the shell. While the kinematical information from the observed shells is consistent with theoretical hydrodynamical models, the observed double-shell structure is not easy to reconcile with the models, as it is unclear whether CO would be destroyed at the centre of a shell.

Based on the CO observations only, it is difficult to derive a comprehensive understanding of the formation and evolution of the shells. Full 3-dimensional hydrodynamical models, coupled with radiative transfer, will be necessary to gain a full understanding of the shells. 

Additionally, it would be necessary to add observations that can probe the shock conditions and photodissociation properties in the shells. For example, CI has been observed in the shell around R~Scl with APEX~\citep{olofssonetal2015}, and would be a dissociation product of CO. Observations of CI at high resolution towards the detached-shell sources may shed light on the distribution of CO in the shells, and directly provide an answer as to whether CO is dissociated in the shell. Direct observations of the H$_2$ distribution would also provide us with information on the formation and evolution of the shells.

Finally, our observations also show arc structures outside the detached shell around V644~Sco, and emission inside the shell towards DR~Ser. It is not clear how the structures around V644~Sco are created. These structures would have to be explained in future models.
\begin{acknowledgements}
This paper makes use of the following ALMA data: ADS/JAO.ALMA\#2015.1.00007.S,~ADS/JAO.ALMA\#2017.1.00006.S, ADS/JAO.ALMA\#2018.1.00010.S,~ADS/JAO.ALMA\#2019.1.00021.S, and ADS/JAO.ALMA\#2021.1.00313.S. ALMA is a partnership of ESO (representing its member states), NSF (USA) and NINS (Japan), together with NRC (Canada), NSTC and ASIAA (Taiwan), and KASI (Republic of Korea), in cooperation with the Republic of Chile. The Joint ALMA Observatory is operated by ESO, AUI/NRAO and NAOJ. TK acknowledges support from the Swedish Research Council through grant 2019-03777.
\end{acknowledgements}


\bibliographystyle{aa} 
\bibliography{maercker}

\begin{appendix}
\section{Channel maps}
\label{a:channelmaps}
\begin{figure*}[h]
\centering
\includegraphics[width=19cm]{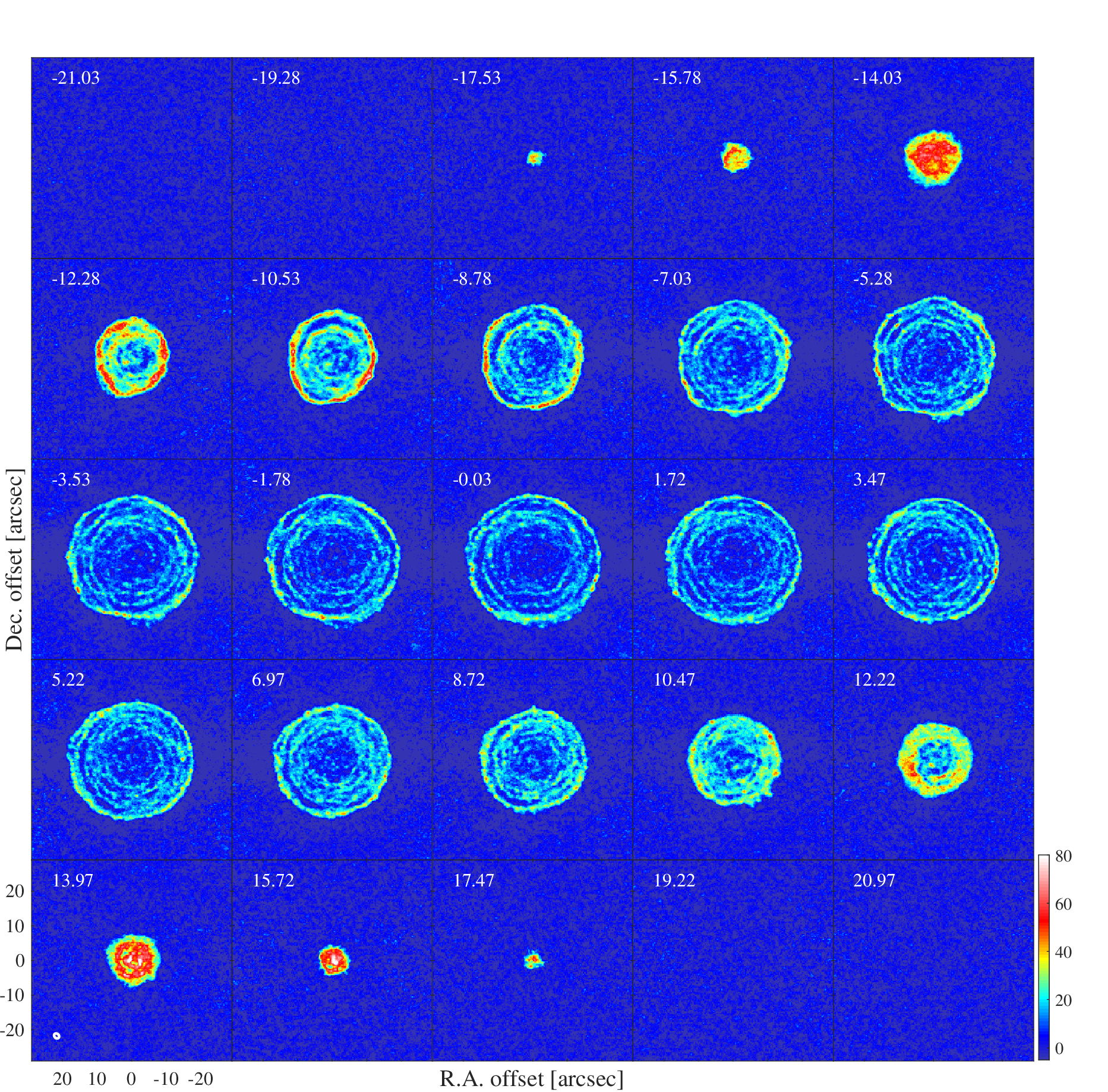} 
\caption{Channel maps of R~Scl in \COone. The colorbar is given in mJy/beam. The numbers in each panel give the velocity in \kms~relative to the \vlsr. The beam is indicated in the lower left panel.}
\label{f:rsclmaps}
\end{figure*}

\begin{figure*}[t]
\centering
\includegraphics[width=19cm]{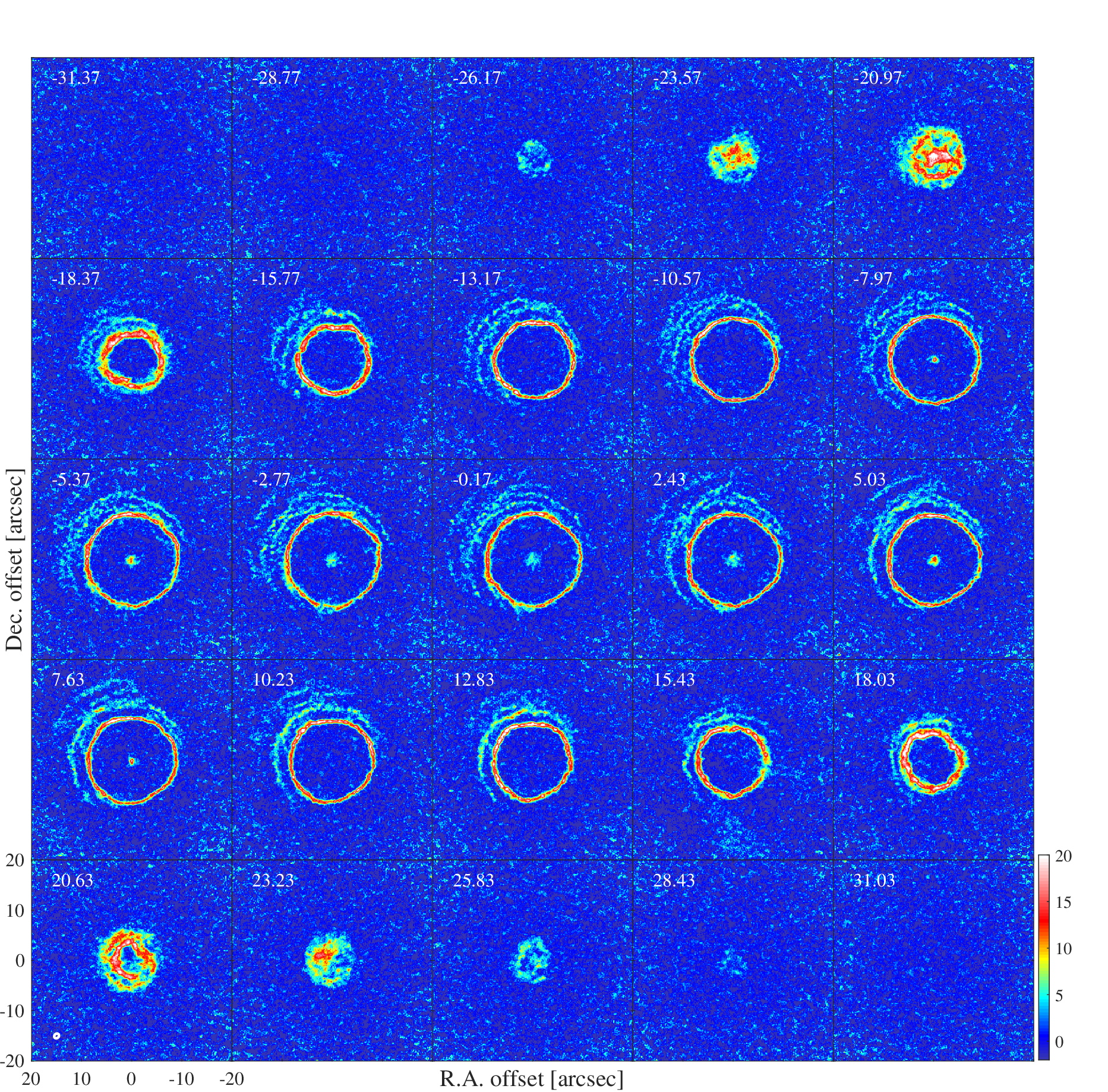} 
\caption{Channel maps of V644~Sco in \COone. The colorbar is given in mJy/beam. The numbers in each panel give the velocity in \kms~relative to the \vlsr. The beam is indicated in the lower left panel.}
\label{f:v644maps}
\end{figure*}

\begin{figure*}[t]
\centering
\includegraphics[width=19cm]{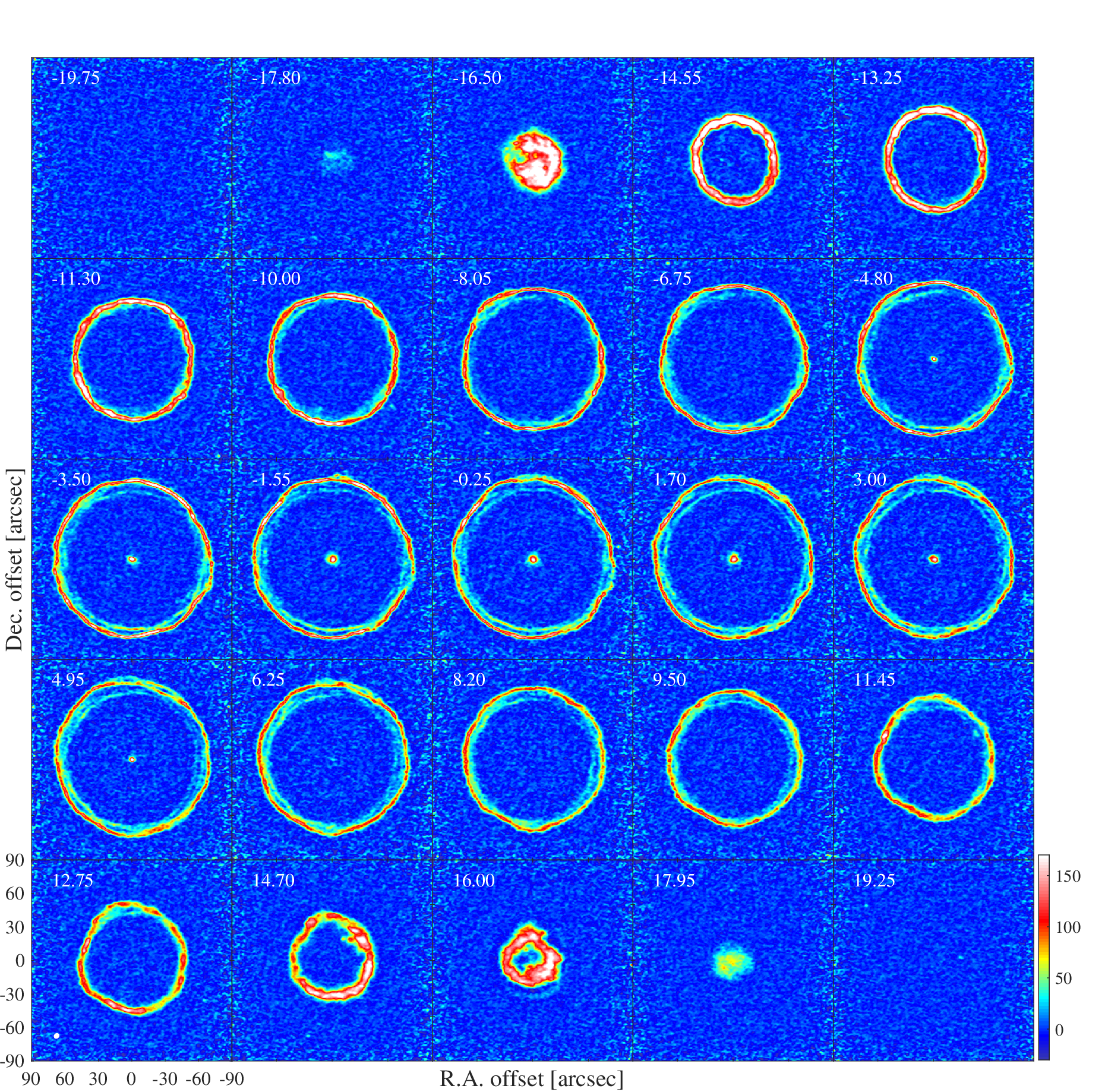} 
\caption{Channel maps of S~Sct in \COone. The colorbar is given in mJy/beam. The numbers in each panel give the velocity in \kms~relative to the \vlsr. The beam is indicated in the lower left panel.}
\label{f:ssctmaps}
\end{figure*}

\begin{figure*}[t]
\centering
\includegraphics[width=19cm]{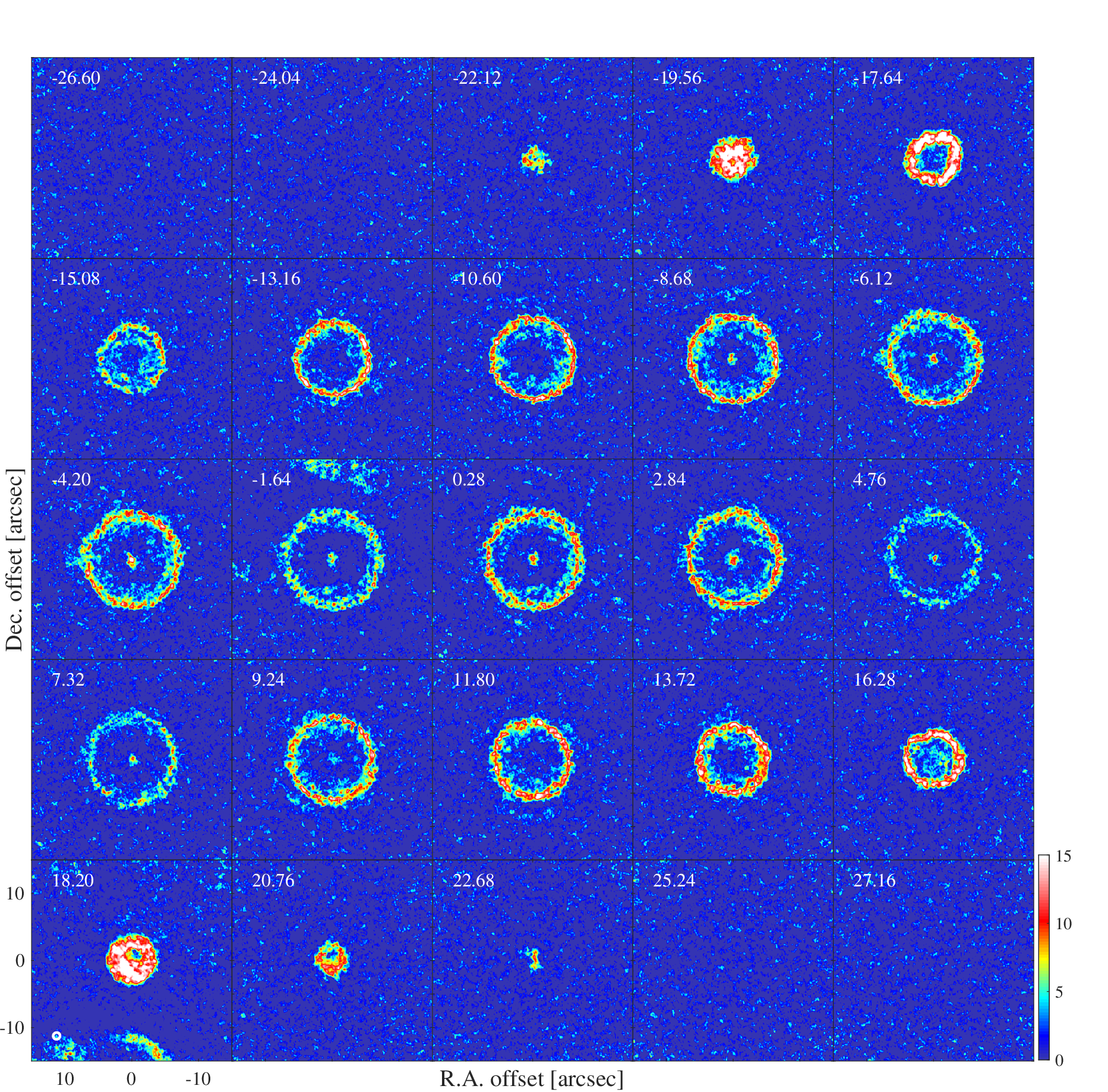} 
\caption{Channel maps of DR~Ser in \COone. The colorbar is given in mJy/beam. The numbers in each panel give the velocity in \kms~relative to the \vlsr.  The beam is indicated in the lower left panel. In some of the frames extended emission can be seen, likely from interstellar emission that is not associated with DR~Ser.}
\label{f:drsermaps}
\end{figure*}

\begin{figure*}[t]
\centering
\includegraphics[width=19cm]{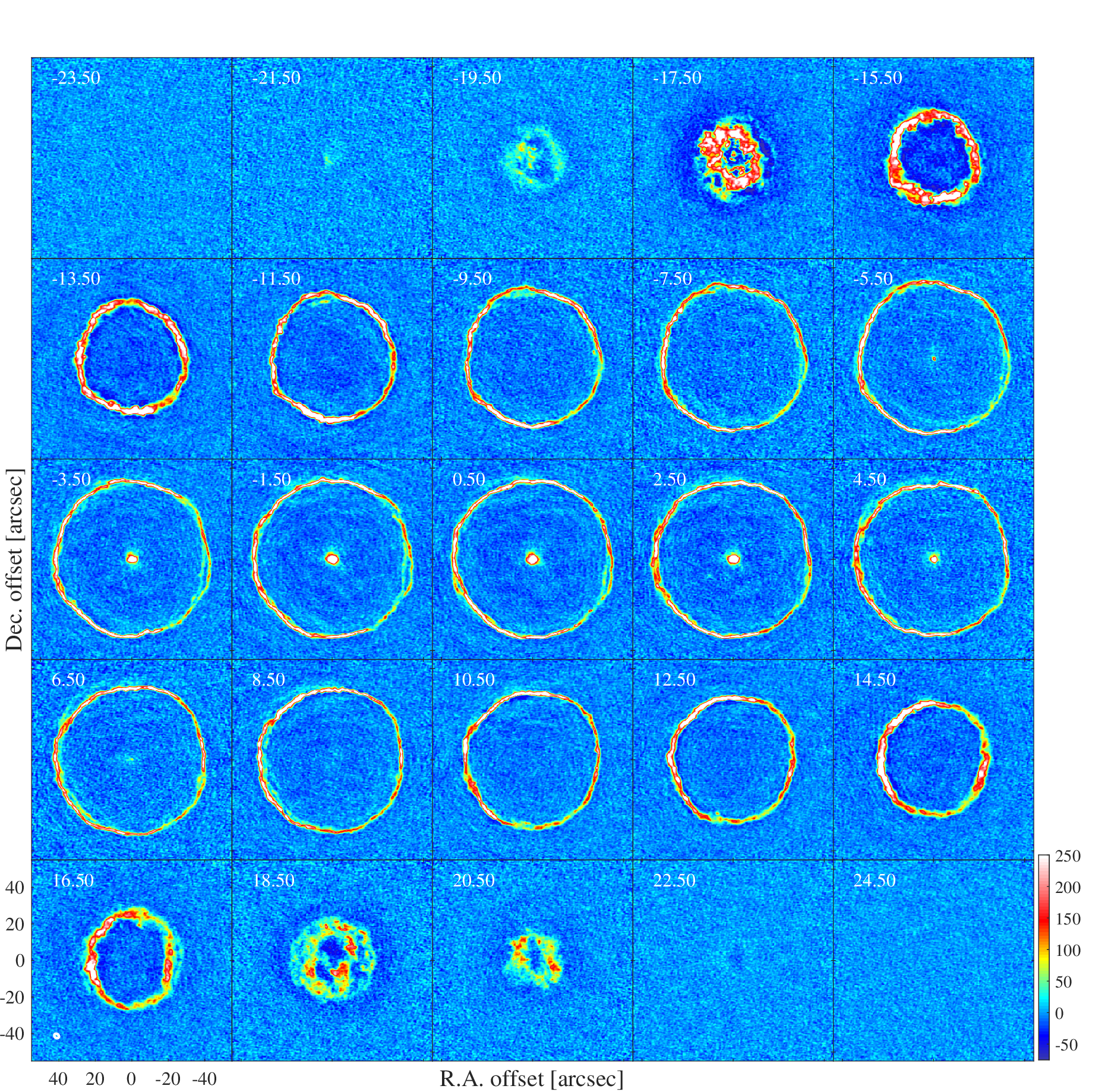} 
\caption{Channel maps of U~Ant in \COtwo. The colorbar is given in mJy/beam. The numbers in each panel give the velocity in \kms~relative to the \vlsr. The beam is indicated in the lower left panel.}
\label{f:uantmaps}
\end{figure*}
\end{appendix}
\end{document}